\makeatletter
\newcommand{\hyper@nopatch@sectioning}{}
\makeatother

\RequirePackage{silence}
\documentclass[aps,prd,preprint,superscriptaddress]{revtex4-2}

\usepackage{array}[=2016-10-06]
\usepackage{graphicx}
\usepackage{dcolumn}
\usepackage{bm}           
\usepackage{cancel}       
\usepackage{amsmath}
\usepackage{amssymb}
\usepackage[utf8]{inputenc}
\usepackage{listings}
\usepackage{url}
\usepackage{xcolor} 
\usepackage{booktabs}
\usepackage[hypertexnames=true]{hyperref} 

\hypersetup{
    colorlinks=true,
    linkcolor=blue,
    citecolor=blue,
    urlcolor=blue
}

\makeatletter
\newcommand{\customlabel}[2]{%
   \protected@write\@auxout{}{%
     \string\newlabel{#1}{{#2}{\thepage}{#2}{#1}{}}%
   }%
   \hypertarget{#1}{}
}
\makeatother

\begin{document}

\title{Neutrino Mass and Mixing Predictions from an Axiomatic \texorpdfstring{$A_5 \rtimes$}{A5 x} CP Topological Manifold}
\author{Enrique Arrieta-Diaz}
 \email{earrieta@unimagdalena.edu.co}
\affiliation{College of Engineering, Universidad del Magdalena, Calle 29H3 No 22 - 01 Santa Marta D.T.C.H., Magdalena, Colombia. Código Postal No. 470004}



\begin{abstract}

This work presents a four-postulate neutrino mass and mixing architecture built on an icosahedral $A_5 \rtimes$ CP flavor symmetry and the three-strand Iwahori--Hecke algebra $\mathcal{H}_3(q)$ of type $A_2$. Matching the discrete curvature of an icosahedral vertex to the six Weyl chambers of the planar $A_2$ complex fixes a single angular defect, $\Delta_{AD} = \pi/18$. In the heavy right-handed Majorana sector, taking the mass operator to be a strictly positive affine combination of the identity and the nearest-neighbor Temperley--Lieb bond sum $B = e_1 + e_2$ on the hyperbolic branch $d = 2\cosh(\Delta_{AD})$ enforces a hierarchical heavy spectrum and an exact, spectral-parameter-independent inverse-mass sum rule. Integrating out the heavy states via a flavor-isotropic Type-I Seesaw mechanism calibrated solely to the empirical mass-squared splittings selects the normal ordering and predicts $m_1 \approx 3.65$ meV and $\sum_i m_i \approx 63.4$ meV. In the mixing sector, preserving $\mathbb{Z}_5$ in the charged-lepton basis and $\mathbb{Z}_2 \times$ CP in the symmetric neutrino parent with $(1,3)$ rotation angle $\Delta_{AD}$ fixes $\theta_{13} = 8.49^\circ$, $\theta_{23} = 45^\circ$, $|\delta_{\mathrm{CP}}| = \pi/2$, and a parent $\theta^{(0)}_{12} = 32.11^\circ$ now disfavored by the latest JUNO data. Including the orthogonal bond-asymmetric operator $e_1 - e_2$ with the post-JUNO rotation postulate $\varepsilon = \Delta_{AD}^2$ preserves $\mu$--$\tau$ reflection invariance while shifting the committed falsification point to $m_{\beta\beta} \approx 1.49$ meV, $m_\beta \approx 9.55$ meV, and $\sin^2\theta_{12} = 0.3104$, to be tested within $[0.302, 0.312]$ at JUNO's 6 year-exposure. Flavor-isotropic Dirac couplings and the hierarchical normal spectrum keep both heavy-threshold and Standard Model radiative corrections negligible.

\end{abstract}


\maketitle

\section{INTRODUCTION}

Despite the advent of high-precision neutrino oscillation data \cite{NOvAT2K2025, T2K2025, NOvA2026, JUNO2026}, the theoretical architecture governing the active mass spectrum and the Pontecorvo--Maki--Nakagawa--Sakata (PMNS) mixing matrix \cite{Pontecorvo1957MA, Pontecorvo1957IBP, Pontecorvo1967, Maki1962, Bilenky1978} remains a profound open question \cite{Altarelli2010}. Current phenomenological models build the mixing matrix, $U^s_{\mathrm{PMNS}}$, as an empirical construct, treating the known mixing angles and mass-squared splittings as continuous inputs extracted from global fits to oscillation data (see Sec.~14 in Ref.~\cite{PDG2024} and Refs.~\cite{nufit2024, NuFit61}) rather than fundamental outputs. Consequently, the field lacks a first-principles mandate for the most critical outstanding observables: the absolute mass scale \cite{Planck2018, DESI2025, KATRIN2025}, the mass hierarchy \cite{JUNO2015}, the Majorana nature of neutrinos \cite{KamLAND_Zen2022}, and the origin \cite{Chen2009,Chen2014} of both the Dirac and Majorana \cite{Bilenky1980,Schechter1980} Charge-Parity-violating phases \cite{Hyper-Kamiokande2018, DUNE2020}.

This work leads to mixing and inverse-mass sum rules, strict parameter correlations, and a constrained absolute mass scale, all of which upcoming high-precision oscillation and cosmological experiments will test.
The architecture is based on the algebra and geometry of an icosahedral $A_5 \rtimes$ CP flavor manifold coupled to the Quantum Yang--Baxter Equation (QYBE) \cite{Yang1967, Baxter1972, Baxter1982, Jimbo1989, Kauffman2002} and the three-strand Iwahori--Hecke algebra $\mathcal{H}_3(q)$, associated with the $A_2$ Coxeter complex \cite{Iwahori1964}, at the heavy Majorana threshold, or Ultraviolet Cutoff \cite{Weinberg1979}, $M_0$. The discrete point group $A_5$ is geometrically isomorphic to the proper rotational symmetries of a regular icosahedron, whose mathematical structures inherently involve the Golden Ratio (GR), $\phi$. To generate fixed, predictable values for the PMNS mixing matrix, Refs.~\cite{Everett2009, Feruglio2011, Ding2012, Feruglio2013} break the icosahedral symmetry to specific residual subgroups in the neutrino and charged-lepton sectors. The QYBE is a fundamental consistency equation that serves as the mathematical foundation for integrable quantum spin chains, quantum groups, and knot invariants, and is applied across a wide variety of physical contexts \cite{Nichita2012}, including particle physics \cite{Artin1947, Faddeev1994, Bilson2005, Arraut2019, Arraut2026}.

The architecture bridges the Electroweak and the Ultraviolet (UV) scales across the heavy Majorana threshold $M_0$. At this scale, three heavy right-handed Majorana neutrino fields, $N_R$ (see App.~\ref{RHMNF}), with mass eigenstates $M_i$, build the Majorana mass matrix $\mathcal{M}_R$, whose symmetry breaking is governed by an icosahedral $A_5 \rtimes$ CP representation coupled to the six-dimensional Iwahori-Hecke algebra $\mathcal{H}_3(q)$. 

This architecture establishes a unified geometric origin for UV mass generation and low-energy flavor mixing by anchoring them to the dual nature of the flavor manifold. While the UV mass spectrum is obtained from the internal algebra of $\mathcal{H}_3(q)$, the low-energy flavor mixing is governed by the manifold's external topology, where the discrete angular defect $\Delta_{AD}$ is directly tested by the reactor angle $\theta_{13}$. Integrating out the heavy states down to the Electroweak scale via the Type-I Seesaw mechanism (see Ch.~6 in Ref. \cite{Giunti2007} and Ref. \cite{Minkowski1977}) yields a normal mass hierarchy (NH) for the active mass spectrum, $m_1 < m_3$, alongside closed-form expressions for the PMNS mixing angles and Charge-Parity (CP) phases. Anchored by $\Delta_{AD}$ and calibrated solely to the two empirical mass-squared splittings: $\Delta m^2_{21}$ and $\Delta m^2_{31}$, the architecture predicts maximal Dirac CP violation: $|\delta_{\mathrm{CP}}| = \pi/2$, maximal atmospheric mixing: $\theta_{23} = 45^\circ$, a reactor angle $\theta_{13} = 8.49^\circ$, a topologically lifted solar angle $\theta_{12} = 33.86^\circ$, and a lightest neutrino mass $m_1 \approx 3.65$ meV, establishing sharp effective mass targets for upcoming neutrinoless double-beta ($0\nu\beta\beta$) decay searches (see Sec.~14.3 in Ref.~\cite{Mohapatra2004} and Refs.~\cite{Wolfenstein1981,Schechter1982,deGouvea2002,Kayser_April_2005,Rodejohann2011}) and cosmological surveys \cite{Lesgourgues2006}.

This work is organized as follows. Sec.~\ref{UHE_Epoch} presents the pristine $A_5$ geometry in the UV regime, $\mu \gg M_0$, where the heavy states can exist on-shell. Sec.~\ref{TBT_epoch} introduces the Iwahori--Hecke and Temperley--Lieb braiding algebra that activates as the running energy scale $\mu$ descends to the heavy Majorana scale, followed by the spectral Baxterization procedure and the Pasquier--Saleur one-magnon eigenbasis in Sec.~\ref{Baxter_sec}. Sec.~\ref{LE_epoch} formulates the single-basis Type-I Seesaw Lagrangian and obtains the heavy Majorana mass eigenvalues. Sec.~\ref{DPT} presents the flavor-basis transformation $\widetilde{U}_R(\Delta_{AD})$ obtained from the residual symmetries $\mathbb{Z}_5$ and $\mathbb{Z}_2 \times$ CP, complemented by App.~\ref{TUMCI}. Sec.~\ref{AIMP} establishes the discrete CP eigenvalues and relative Majorana phases. The PMNS mixing angles, Jarlskog invariant, and astrophysical neutrino telescope flavor ratios are obtained in Sec.~\ref{PMNSMA}, while the implications of latest results on the solar angle are discussed in Sec.~\ref{App_Extension}. The active neutrino mass spectrum and the exact inverse-mass sum rule are presented in Sec.~\ref{AFAM}. The effective neutrino masses $m_{\beta \beta}$ and $m_{\beta}$ are calculated in Sec.~\ref{EffM}. Finally, the conclusions and summary tables of postulates and predictions are presented in Sec.~\ref{conclusion}. The properties of right-handed Majorana neutrino fields are summarized in App.~\ref{RHMNF} and the radiative stability of the architecture across heavy thresholds and down to the Electroweak scale is proven in App.~\ref{RGE_I}.

\section{ASYMPTOTIC \texorpdfstring{$A_5$}{A5} FLAVOR GEOMETRY}
\label{UHE_Epoch}

Standard discrete flavor models \cite{Everett2009, Feruglio2011, Ding2012} treat the observable PMNS mixing angles as kinematic projections of a broken discrete point group such as $A_5$, in which the left-handed lepton doublets are assigned to the three-dimensional irreducible triplet representation $\mathbf{3}$. Specifically, an unbroken Klein four-group $V_4 = \mathbb{Z}_2 \times \mathbb{Z}_2 \subset A_5$ alignment in the neutrino sector against a cyclic $\mathbb{Z}_5$ alignment in the charged-lepton sector yields the GR mixing matrix, $U_{\mathrm{GR}}$, which natively predicts $\theta^{(0)}_{13} = 0$. Nevertheless, precision reactor oscillation experiments definitively establish that $\theta_{13} \neq 0$ \cite{DayaBay2012,DayaBay2022}, while $\tan(\theta_{12}) \sim 1/\phi$ and $\cos(\theta_{23}) \sim 1/\sqrt{2}$. Hence, $U_{\mathrm{GR}}$ cannot by itself represent the complete mixing matrix $U^s_{\mathrm{PMNS}}$. This architecture factorizes $U_{\mathrm{PMNS}}$ into the flavor-symmetric GR baseline perturbed by a residual $\mathbb{Z}_2 \times$ CP unitary boundary twist \cite{Li2015, DiIura2015, Ballett2015}, replacing empirical free angles with closed-form geometric relations.

Above the UV boundary, where the energy scale $\mu \gg M_0$, the internal flavor space is governed by the unbroken icosahedral point group $A_5 \subset SO(3)$, yielding the unperturbed baseline mixing matrix $U_{\mathrm{GR}}$ in the irreducible triplet representation $S^2 = \mathcal{T}^3 = (S\mathcal{T})^5 = I$, where $S$ is a $\mathbb{Z}_2$ involution and $\mathcal{T}$ and $S\mathcal{T}$ generate $\mathbb{Z}_3$ and $\mathbb{Z}_5$ cyclic subgroups, respectively (see Refs.~\cite{Everett2009, Ding2012}). The five-fold cycle phase $q_H = e^{i \pi/5}$ defines the quantum dimension $[2]_{q_H}$ of the Fibonacci anyon \cite{Jones1987, Nayak2008}
\begin{equation} 
    \nonumber
    d_H = [2]_{q_H} = q_H + q^{-1}_H = 2 \cos \left( \frac{\pi}{5} \right) = \frac{1 + \sqrt{5}}{2} \equiv \phi,
\end{equation}
satisfying $d^2_H - d_H - 1 = 0$ and $2 + \phi = \sqrt{5}\phi$. The non-zero entries of the electron row of $U_{\mathrm{GR}}$, $(\phi, 1)/\sqrt{2+\phi} = (0.8507, 0.5257)$, match up to a sign the second row of the unitary Fibonacci modular $S$-matrix (see Ref.~\cite{Nayak2008})
\begin{equation}
    \nonumber
    S_{\mathrm{Fib}} = \frac{1}{\sqrt{2 + \phi}} \begin{pmatrix} 1 & \phi \\ \phi & -1 \end{pmatrix},
\end{equation}
immediately enforcing the unperturbed GR solar baseline $\tan(\theta^{(0)}_{12}) = 1/\phi$
\begin{equation}
    \label{P_U_GR}
    U_{\mathrm{GR}} = \begin{pmatrix}
            \sqrt{\frac{\phi}{\sqrt{5}}} & \sqrt{\frac{1}{\sqrt{5}\phi}} & 0\\
            -\sqrt{\frac{1}{2\sqrt{5}\phi}} & \sqrt{\frac{\phi}{2\sqrt{5}}} & \frac{1}{\sqrt{2}}\\
            \sqrt{\frac{1}{2\sqrt{5}\phi}} & -\sqrt{\frac{\phi}{2\sqrt{5}}} & \frac{1}{\sqrt{2}}\\
        \end{pmatrix}.
\end{equation}
In Eq.~(\ref{P_U_GR}), $U_{\mathrm{GR}} = \mathcal{P} \cdot U^{\mathrm{Everett}}_{\mathrm{GR}} \cdot \mathcal{P}$ with $\mathcal{P} = \mathrm{diag}(1,1,-1)$ (see Ref.~\cite{Everett2009} for the full derivation).

At this asymptotic scale, the active left-handed lepton fields remain massless while the heavy right-handed Majorana sector possesses the primordial mass scale $M_0$. In the undeformed, unbraided state, the Klein four-group eigenbasis is characterized by the discrete involution operator $S = \mathrm{diag}(s_1, s_2, s_3)$, with minimal polynomial $(S-I)(S+I) = 0$ and eigenvalues $s_1 = +1$ and $s_2 = s_3 = -1$. In contrast to the external five-fold icosahedral phase $q_H = e^{i\pi/5}$, the internal three-generation braid algebra $\mathcal{H}_3(q)$ is governed by an independent deformation exponent $\Delta$. For the massless gauge-invariant Standard Model (SM) fields, this deformation vanishes: $\Delta \to \Delta_0 = 0, \, q_0 = e^{\Delta_0} = 1$, thus, $d_0 = 2$, where $d_0$ is the quantum dimension $[2]_{q_0}$. However, at the heavy Majorana scale, $\mu \sim M_0$, the lepton-number-violating (see App. \ref{RHMNF}) heavy Majorana mass matrix is structurally governed by the deformed Iwahori--Hecke algebra $\mathcal{H}_3(q'_{AD})$ with $q'_{AD} = e^{\Delta_{AD}} \neq 1$. This topological boundary condition lifts the three-fold mass degeneracy and generates the exact heavy Majorana mass spectrum.

\section{IWAHORI-HECKE BRAIDING AND THE QYBE}
\label{TBT_epoch}

Fixing the deformation parameter at the heavy Majorana scale requires matching the algebraic braiding of $\mathcal{H}_3(q'_{AD})$ to the discrete polyhedral curvature of the $A_5$ icosahedral vertex. When the scale invariance of the asymptotic $q_0 = 1$ limit is broken at $\mu \sim M_0$, pulling the discrete Gauss--Bonnet curvature concentrated at an icosahedral vertex \cite{Banchoff1967, Coxeter1973} back to the six fundamental Weyl chambers of the flat $A_2$ tangent complex fixes an elementary quantum of angular defect, $\Delta_{AD} = \pi/18$ (see Sec.~\ref{DPT} for the detailed geometric argument). Through the Baxterization procedure (see Sec. \ref{Baxter_sec}) governed by the QYBE, the internal three-generation mass operator is resolved into a self-consistent network of pairwise nearest-neighbor crossings. Following canonical braid group formalisms (see Ref.~\cite{Jones1987}), an n $ = 3$ strand space requires n $ - 1 = 2$ adjacent crossing generators, $S_1$ and $S_2$, representing the $(1,2)$ and $(2,3)$ adjacent-strand crossings, respectively. If the fundamental braid generators were to act as flat, trivial permutations satisfying $S_i^2 = I$, an over-crossing would pass transparently through itself to become an under-crossing, untangling the braid without lifting the three-fold mass degeneracy. In the representation theory of Iwahori--Hecke and Temperley--Lieb algebras (see Refs.~\cite{Temperley1971, Baxter1982, Chari1994}), lifting the mass degeneracy with real eigenvalues drives the deformation parameter off the critical point $q_0 = 1, \, d_0 = 2$, onto the real positive hyperbolic branch $q'_{AD} = e^{\Delta_{AD}} \neq 1$. Embedding the elementary $2 \times 2$ crossing blocks into the $3 \times 3$ auxiliary open-chain strand representation $\mathcal{H}_{\mathrm{strand}}$ yields
\begin{equation}
    \nonumber
    \begin{split}
        S_1 &= \begin{pmatrix} 0 & 1 & 0 \\ 1 & q'_{AD} - q'^{-1}_{AD} & 0 \\ 0 & 0 & q'_{AD} \end{pmatrix}, \\
        S_2 &= \begin{pmatrix} q'_{AD} & 0 & 0 \\ 0 & 0 & 1 \\ 0 & 1 & q'_{AD} - q'^{-1}_{AD} \end{pmatrix},
    \end{split}
\end{equation}
where the generators are asymmetric $S_1 \neq S_2$. In accordance with the underlying Coxeter group algebra (see Chs.~5 and 7 in Ref.~\cite{Humphreys1990}), both generators are non-commutative. The quadratic, orientation-dependent Hecke braid products $S_1 S_2 \neq S_2 S_1$, take the explicit forms
\begin{equation}
    \nonumber
    \begin{split}
        S_1 S_2 &= \begin{pmatrix}
        0 & 0 & 1 \\
        q'_{AD} & 0 &  q'_{AD} - q'^{-1}_{AD} \\
        0 & q'_{AD} & q'^2_{AD} - 1 \\
        \end{pmatrix}, \\
        S_2 S_1 &= \begin{pmatrix}
        0 & q'_{AD} & 0 \\
        0 & 0 & q'_{AD} \\
        1 & q'_{AD} - q'^{-1}_{AD} & q'^2_{AD} - 1 \\
        \end{pmatrix},
    \end{split}
\end{equation}
and the maximal Garside half-twist, $i.e.$, the longest Coxeter--Hecke word, which satisfies the fundamental braid relation (see Ch.~7 in Ref.~\cite{Kauffman1991}), is
\begin{equation}
    \nonumber
    \begin{split}
        S_1 S_2 S_1 &= S_{\max} = S_2 S_1 S_2, \\
        S_{\max} &= \begin{pmatrix}
        0 & 0 & q'_{AD} \\
        0 & q'_{AD} & q'^{2}_{AD} - 1\\
        q'_{AD} & q'^{2}_{AD} - 1 & q'^3_{AD} - q'_{AD} \\
        \end{pmatrix}.
    \end{split}
\end{equation}

In a $q$-deformed manifold, the continuous spectral anisotropy is governed by the Temper\-ley\--Lieb projection generators, $e_i$ (see Ref.~\cite{Temperley1971}), which act as localized nearest-neighbor loop contractions of adjacent strands. Combining the uncrossed propagation with this pairwise Temper\-ley\--Lieb coupling defines the Hecke braid operators via the Jones construction (see Chs.~3 and 4 in Ref.~\cite{Kauffman1991} and Ref.~\cite{Jones1987})
\begin{equation}
    \label{Jones_C}
    S_i = q'_{AD} I - e_i.
\end{equation}
Eq. (\ref{Jones_C}) demonstrates \cite{Kauffman1987} that $S_i$ is a linear superposition of the uncrossed identity channel, $q'_{AD}I$, and the localized Temperley--Lieb projector, $e_i$.

In an integrable quantum chain, the nearest-neighbor Hamiltonian (see Refs.~\cite{Temperley1971, Baxter1982, Pasquier1990}) is constructed as the symmetric sum of the localized Temperley--Lieb generators across adjacent bonds
\begin{equation}
    \label{B_operator_sum}
    B = e_1 + e_2 =  \begin{pmatrix}
        q'_{AD} & -1 & 0 \\
        -1 & d'_{AD} & - 1\\
        0 & -1 & q'^{-1}_{AD} \\
        \end{pmatrix}.
\end{equation}
In the Temperley--Lieb algebra $TL_3(d'_{AD})$, the projection operators satisfy two defining relations (see Chs.~2 and 4 in Ref.~\cite{Kauffman1991} and Ref.~\cite{Temperley1971}):
\begin{itemize}
    \item \textbf{The Entanglement (Braid) Relations}: For adjacent operators, $e_1 e_2 e_1 = e_1$ and $e_2 e_1 e_2 = e_2$.
    \item \textbf{The Loop Weight}: $e^2_i = d'_{AD} e_i$.
\end{itemize}
Here $d'_{AD}$ (see Sec.~10.2 in Ref.~\cite{Chari1994}) is the hyperbolic pairwise quantum dimension $[2]_{q'_{AD}}$
\begin{equation}
    \label{loop_weight}
    \begin{split}
        d'_{AD} &= q'_{AD} + q'^{-1}_{AD} \\
        &= e^{\Delta_{AD}} + e^{-\Delta_{AD}} = 2\cosh(\Delta_{AD}).
    \end{split}
\end{equation}

The basis-independent minimal polynomial of $B$ follows directly from its algebraic powers. Evaluating $B^2$ yields
\begin{equation}
    \label{B_squared}
    B^2 = (e_1 + e_2)^2 = d'_{AD} B + (e_1 e_2 + e_2 e_1),
\end{equation}
and $B^3$ is
\begin{equation}
    \label{B_third}
    \begin{split}
        B^3 &= B(d'_{AD} B + e_1 e_2 + e_2 e_1) \\
        &= d'_{AD} B^2 + (e_1 + e_2)(e_1 e_2 + e_2 e_1).
    \end{split}
\end{equation}
Expanding the second term in Eq.~(\ref{B_third}) via the Temperley--Lieb entanglement relations gives
\begin{equation}
    \nonumber
    (e_1 + e_2)(e_1 e_2 + e_2 e_1) = B + d'_{AD} (e_1 e_2 + e_2 e_1).
\end{equation}
Substituting Eq.~(\ref{B_squared}) into Eq.~(\ref{B_third})
\begin{equation}
    \nonumber
    \begin{split}
        B^3 &= d'_{AD} B^2 + B + d'_{AD} (B^2 - d'_{AD} B) \\
        &= 2 d'_{AD} B^2 + (1 - d'^2_{AD}) B,
    \end{split}
\end{equation}
results in the exact cubic minimal polynomial for the coupled Temperley--Lieb Hamiltonian
\begin{equation}
    \nonumber
    B^3 - 2 d'_{AD} B^2 + (d'^2_{AD} - 1) B = B\left[(B - d'_{AD}I)^2 - I\right] = 0.
\end{equation}
Factoring the characteristic polynomial
\begin{equation} \nonumber
    b [ b^2 - 2 d'_{AD} b + (d'^2_{AD} - 1) ] = 0,
\end{equation}
fixes the three exact algebraic roots $b_i$
\begin{equation}
    \label{The_b_i}
    b_1 = d'_{AD} + 1, \qquad b_2 = d'_{AD} - 1, \qquad b_3 = 0.
\end{equation}

\subsection{Spectral Baxterization of the Mass Operator}
\label{Baxter_sec}

To transition the static braid generators into a continuous spectral architecture, the Hecke generators are subjected to Baxterization. Introducing the continuous Yang--Baxter spectral parameter $u$ via the multiplicative parameter $x = e^{-u}$ generates two continuous $R_i(x)$ matrices \cite{Jones1990}
\begin{equation}
    \nonumber
    R_i(x) = \rho \left (x S_i - x^{-1} S_i^{-1} \right),
\end{equation}
where $\rho$ is a normalization factor. Since the $S_i$ fulfill the Hecke quadratic condition
\begin{equation}
    \nonumber
    S^2_i = (q'_{AD} - q'^{-1}_{AD}) S_i + I,
\end{equation}
the inverse relation $S_i^{-1} = S_i - (q'_{AD} - q'^{-1}_{AD})I$ brings $R_i(x)$ into the linear form
\begin{equation}
    \label{The_R_i}
    R_i(x) = \rho \left [ (x - x^{-1}) S_i + x^{-1} (q'_{AD} - q'^{-1}_{AD}) I \right ],
\end{equation}
which satisfies the spectral QYBE (see Sec.~9.6 in Ref.~\cite{Baxter1982} and Sec.~12.2 in Ref.~\cite{Chari1994})
\begin{equation} \nonumber
    R_1(u)R_2(u+v)R_1(v) = R_2(v)R_1(u+v)R_2(u).
\end{equation}
Applying the Jones construction of Eq.~(\ref{Jones_C}) to Eq.~(\ref{The_R_i}) yields the local Temperley--Lieb decomposition
\begin{equation}
    \nonumber
    \begin{split}
        R_i(x) &= \rho \left [ \left (x - x^{-1} \right) (q'_{AD} I - e_i) + x^{-1} \left (q'_{AD} - q'^{-1}_{AD} \right) I \right ] \\
        &= \rho \left [ \left (e^{-u} e^{\Delta_{AD}} - e^{u} e^{-\Delta_{AD}}  \right ) I  - \left (e^{-u} - e^{u} \right) e_i \right ] \\
        &= 2 \rho \left [ \sinh(\Delta_{AD} - u) I  +  \sinh(u) e_i \right ]. 
    \end{split}
\end{equation}
Enforcing the normalization condition $R_i(x=1) = I$ in the unbraided limit $u = 0$ fixes $\rho$
\begin{equation}
    \nonumber
    \rho = \big(2 \sinh(\Delta_{AD})\big)^{-1} = \left ( q'_{AD} - q'^{-1}_{AD} \right )^{-1}.
\end{equation}

In the five-dimensional Temper\-ley\--Lieb algebra $TL_3(d'_{AD}) = \mathrm{span}\{I, e_1, e_2, e_1 e_2, e_2 e_1\}$, the subalgebra invariant under the canonical generator-exchange automorphism  $\sigma: e_1 \leftrightarrow e_2$ is three-dimensional, spanned by $\{I, \, B, \, B^2\}$ where
\begin{equation}
    \nonumber
    e_1e_2 + e_2e_1 = B^2 - d'_{AD}B,
\end{equation}
by Eq.~(\ref{B_squared}). Consequently, any $\sigma$-symmetric operator $f(B)$ commutes with $B$ and shares its eigenvectors.

In this architecture, the continuous spectral parameter $u$ is fitted to the empirical mass-squared splittings, and the allowable mass operators are restricted to the mutually commuting conserved charges of the $R$-matrix. The fundamental nearest-neighbor Hamiltonian is generated explicitly by the first-order logarithmic derivative of the QYBE spectral matrices at $u = 0$ (see Ref.~\cite{Baxter1982})
\begin{equation} \nonumber
    \begin{split}
        \sum_{i=1}^2\left.\frac{d}{du}R_i(u)\right|_{u=0} &= -2\coth(\Delta_{AD})I + \sinh^{-1}(\Delta_{AD})B\\
        &\quad  \in \mathrm{span}\{I, B\}.
    \end{split}
\end{equation}
Within the three-dimensional subspace $\mathrm{span}\{I, B, B^2\}$ of operators commuting with $B$, the third postulate (P3) of the architecture: the positive hyperbolic affine mass operator, selects the hyperbolic loop weight of Eq.~(\ref{loop_weight}), restricts the heavy Majorana mass operator to a first-order affine combination of the identity $I$ and the nearest-neighbor bond sum $B$ of Eq.~(\ref{B_operator_sum}), and confines the spectral parameter to the positive domain (PD): $0 < u < \Delta_{AD}$. On this interval, $\sinh(\Delta_{AD}-u)$ and $\sinh(u)$ are strictly positive spectral coefficients for $I$ and $B$, respectively, with their ratio $\sinh(u)/\sinh(\Delta_{AD}-u)$ increasing monotonically from $0$ to $+\infty$. At finite $u$, this corresponds to summing the local two-body bond contributions
\begin{equation} \nonumber
    \delta R_i(u) = R_i(u) - \sinh^{-1}(\Delta_{AD})\sinh(\Delta_{AD}-u)I,
\end{equation}
across both adjacent interfaces while retaining a single uncrossed identity channel
\begin{equation}
    \nonumber
    R(u) = R_1(u) + R_2(u) - \frac{\sinh(\Delta_{AD} - u)}{\sinh(\Delta_{AD})} I.
\end{equation}
This defines the global $3 \times 3$ affine bond-sum operator in $\mathrm{span}\{I, B\}$
\begin{equation}
    \label{Final_R}
    R(u) = \sinh^{-1}(\Delta_{AD}) \Big[ \sinh(\Delta_{AD} - u) I + \sinh(u) B \Big],
\end{equation}
whose seesaw inverse $\mathrm{m}_\nu \propto R^{-1}(u)$, in Sec.~\ref{AFAM}, provides the inverse affine structure governing the light active neutrino mass matrix.

In the coordinate representation of Eq.~(\ref{B_operator_sum}), the $3 \times 3$ matrix $B$ is identical to the one-magnon sector of the three-site, open-boundary $U_q(\mathfrak{sl}_2)$-invariant XXZ quantum chain of Pasquier and Saleur (see Ref.~\cite{Pasquier1990}), where the diagonal boundary entries $q'_{AD}$ and $q'^{-1}_{AD}$ represent the boundary fields required to preserve exact quantum-group invariance. Since Eq.~(\ref{B_operator_sum}) is written in this auxiliary open-chain site basis $\mathcal{H}_{\mathrm{strand}}$, transitioning to the intrinsic spectral eigenbasis $\hat{\mathcal{H}}$ of the Temperley--Lieb quotient diagonalizes $B$ via its orthogonal eigenvector matrix $O$: $O^T B O = \hat{B} = \mathrm{diag}(b_1, b_2, b_3)$. While the affine operator $B$ governs the discrete mass spectrum in this topological site basis, it does not trivially align with the physical lepton flavor basis. The physical flavor basis is defined by the diagonal charged-lepton sector, and is structurally misaligned from the pure topological mass basis by the intrinsic geometric defect of the $A_5$ manifold. This fundamental misalignment is quantified by the Weyl-averaged topological strain postulated in Sec.~\ref{DPT}, which acts as the physical unitary rotation $\widetilde{U}_R(\Delta_{AD})$ in Sec.~\ref{LE_epoch} required to map the abstract chain into the physical weak interaction basis. Thus, the algebra's sole input to the heavy mass sector is the one-magnon eigenvalue spectrum $\{b_1, b_2, b_3\} = \{d'_{AD} + 1, \, d'_{AD} - 1, \, 0\}$, where $b_1$ and $b_2$ span the two-dimensional loop-coupled module $\mathcal{V}_2$ and $b_3 = 0$ spans the one-dimensional $U_q(\mathfrak{sl}_2)$-symmetric trivial module $\mathcal{V}_1$.

In this spectral eigenbasis $\hat{\mathcal{H}} = \mathcal{V}_2 \oplus \mathcal{V}_1$, the bond-exchange automorphism $\sigma$: ($e_1 \leftrightarrow e_2$) acts non-trivially only within the two-dimensional loop-coupled module $\mathcal{V}_2$, locking the solar pair to opposite bond-exchange parities: $s_1 = +1, s_2 = -1$, while leaving the singlet $|b_3\rangle \in \mathcal{V}_1$ invariant with a twofold sign freedom: $s_3 = \pm 1$. 

Scaling the diagonal spectral weight operator
\begin{equation} \nonumber
    \hat{R}(u) = O^T R(u) O = \mathrm{diag}(l_1(u), l_2(u), l_3(u)),
\end{equation}
by the primordial scale $M_0$ and $S$ defines the signed heavy Majorana spectral mass operator
\begin{equation}
    \label{M_R_final}
    \begin{split}
        \Lambda_R(u) &= M_0 S \, \hat{R}(u) \\
        &= \frac{M_0}{\sinh(\Delta_{AD})} S \Big[\sinh(\Delta_{AD} - u)I + \sinh(u)\hat{B}\Big],
    \end{split}
\end{equation}
which recovers the unbraided involution spectrum $\Lambda_R(0) = M_0 S$ in the $u \to 0$ limit. 

\section{TYPE-I SEESAW MECHANISM DECOUPLING AND HEAVY SPECTRUM}
\label{LE_epoch}

To anchor this algebraic structure in an explicit effective field theory, both the SM left-handed lepton doublets $L$ and the heavy right-handed Majorana singlets $N^{\mathrm{(fl)}}_R$ in the flavor basis are assigned to the three-dimensional irreducible triplet representation $\mathbf{3}$ of the icosahedral group $A_5 \subset SO(3)$ (see Refs.~\cite{Everett2009, Ding2012}), where Schur's lemma enforces a flavor-isotropic Dirac Yukawa coupling $y_D \overline{L} I_{3\times 3} \tilde{H} N^{\mathrm{(fl)}}_R$. Moreover, the internal Temperley--Lieb algebra $TL_3(d'_{AD})$ fixes the heavy eigenvalue spectrum $\Lambda_R(u)$ in the spectral eigenbasis $\hat{N}_R = (\hat{N}_{R,1}, \hat{N}_{R,2}, \hat{N}_{R,3})^T$. 

However, the relative orientation between $\hat{N}_R$ and the charged-lepton flavor basis $N^{\mathrm{(fl)}}_R = \widetilde{U}_R(\Delta_{AD}) \hat{N}_R$ is not determined by the algebra. Instead, $\widetilde{U}_R(\Delta_{AD})$ is governed by the residual symmetries of $A_5 \rtimes$ CP (see Refs.~\cite{Ding2012, Feruglio2013}): whereas the charged-lepton mass basis is defined as the eigenbasis of the cyclic $\mathbb{Z}_5$ generator $\mathcal{T}$, which is separated from the unbroken Klein four-group $V_4 = \mathbb{Z}_2 \times \mathbb{Z}_2$ eigenbasis by $U_{\mathrm{GR}}$ (see Refs.~\cite{Everett2009, Ding2012}), breaking $V_4$ in the Majorana sector to a single residual $\mathbb{Z}_2 \times$ CP symmetry that preserves the second column of $U_{\mathrm{GR}}$ restricts $\widetilde{U}_R(\theta)$ to a one-parameter family of $(1,3)$-eigenplane rotations up to a single angle $\theta$ and discrete CP eigenvalues $s_i \in \{+1, -1\}$, with: $i = 1, 2, 3$. In Sec.~\ref{DPT}, $\widetilde{U}_R(\Delta_{AD})$ is constructed explicitly and $|\theta| = \Delta_{AD}$ is fixed by matching to the six-chamber $A_2$ Coxeter complex. Above the heavy Majorana scale, the renormalizable Type-I Seesaw Lagrangian (see Ref.~\cite{Minkowski1977} and Ch.~6 in Ref.~\cite{Giunti2007}) takes the single-basis form
\begin{equation}
    \nonumber
    \begin{split}
        -\mathcal{L}_{\mathrm{seesaw}} &= y_D \, \overline{L} \, \widetilde{U}_R(\Delta_{AD}) \tilde{H} \hat{N}_R \\
        &\quad + \frac{1}{2} \overline{\hat{N}^c_R} \, \Lambda_R(u) \, \hat{N}_R + \mathrm{h.c.} \\
        &= y_D \, \overline{L} \, I_{3\times 3} \tilde{H} N^{\mathrm{(fl)}}_R \\
        &\quad + \frac{1}{2} \overline{\left(N^{\mathrm{(fl)}}_R\right)^c} \, \mathcal{M}^{\mathrm{(fl)}}_R(u) \, N^{\mathrm{(fl)}}_R + \mathrm{h.c.},
    \end{split}
\end{equation}
where $\tilde{H} = i\sigma_2 H^*$ is the conjugate Higgs doublet. In the flavor basis $(L, N^{\mathrm{(fl)}}_R)$, spontaneous Electroweak symmetry breaking \cite{Glashow1961, Weinberg1967, Salam1968, Englert1964, Higgs1964, Guralnik1964, tHooft1972} yields the fla-vor-isotropic Dirac mass matrix $M_D = m_D I_{3\times 3}$ with universal Dirac mass scale $m_D = y_D v_{\mathrm{EW}}/\sqrt{2}$, where $v_{\mathrm{EW}} = (\sqrt{2}G_F)^{-1/2} = 246.22$ GeV is the SM Higgs vacuum expectation value (see Sec.~10.1 in Ref.~\cite{PDG2024}).

Because the symmetric affine bond-sum operator $B$ in Eq.~(\ref{B_operator_sum}) takes the form of a nearest-neighbor tridiagonal matrix in the site basis, this structure hints at a geometric origin for the heavy Majorana scale and invites a field-theoretic interpretation in which three right-handed Majorana singlets reside on a three-site chain, such as a deconstructed extra dimension~\cite{ArkaniHamed2001} or a flavor-clockwork lattice~\cite{Giudice2017}, with adjacent sites coupled by nearest-neighbor mass insertions. Extending this three-site picture to a full flavor theory requires an explicit flavon scalar potential that breaks $A_5 \rtimes$ CP selectively to $\mathbb{Z}_5$ in the charged-lepton sector and to $\mathbb{Z}_2 \times \mathrm{CP}$ in the Majorana sector, with bond asymmetry softly breaking $\mathbb{Z}_2$ while preserving residual $\mu$--$\tau$ reflection $X$, and stabilizes vacuum alignments corresponding to $|\theta| = \Delta_{AD}$ and $\varepsilon = \Delta_{AD}^2$ (see Secs.~\ref{DPT} and~\ref{App_Extension}).

In addition, a UV completion must explain why the Dirac Yukawa matrix remains flavor-isotropic,
\begin{equation}
    \nonumber
    Y_D = y_D I_{3\times 3},
\end{equation}
at leading order despite $\mathcal{O}(1)$ $A_5$-breaking in the Majorana sector, e.g., through auxiliary shaping symmetries (see Refs.~\cite{Feruglio2011, Ding2012}) or sequestered brane localization (see Refs.~\cite{Altarelli2005, Altarelli2010}). Notably, exact isotropy enforces $Y_D^\dagger Y_D = |y_D|^2 I_{3\times 3}$, causing the CP asymmetries $\epsilon_i$ of standard thermal leptogenesis~\cite{Fukugita1986, Davidson2008} to vanish identically at leading order; realizing leptogenesis within this architecture would therefore require small, higher-dimensional flavon corrections to $Y_D$~\cite{Jenkins2008, Bertuzzo2009}.

Descending in effective energy scale to the Dirac mass scale, $M_D$ and $\mathcal{M}^{\mathrm{(fl)}}_R(u)$ combine within the canonical $6 \times 6$ Type-I seesaw matrix in the flavor basis $(L, N^{\mathrm{(fl)}}_R)$ (see Ref.~\cite{Minkowski1977} and Ch.~6 in Ref.~\cite{Giunti2007})
\begin{equation} \nonumber
    M_{\nu} = \begin{pmatrix}
                  0_{3 \times 3} & M_D \\
                  M^T_D & \mathcal{M}^{\mathrm{(fl)}}_R (u)
              \end{pmatrix},
\end{equation}
where the null sub-block $0_{3 \times 3}$ reflects the absence of direct tree-level left-handed Majorana mass terms. Using the hyperbolic bond-sum eigenvalues from Eq.~(\ref{M_R_final}), the one-magnon roots $b_i$ from Eq.~(\ref{The_b_i}), and the discrete CP eigenvalues $s_i$, explicitly fixed in Eq.(\ref{d_AD_prime}) below (with $S = \mathrm{diag}(s_1, s_2, s_3)$), the signed diagonal heavy Majorana eigenvalue matrix $\Lambda_R(u)$ takes the form
\begin{equation}
    \label{M_0_eigenvalues}
    \begin{split}
        \Lambda_R(u) &= M_0 S \, \mathrm{diag}\big(l_1 (u), l_2 (u), l_3 (u)\big) \\
        &= M_0 \, \mathrm{diag}\big(s_1 l_1 (u), s_2 l_2 (u), s_3 l_3 (u)\big),
    \end{split}
\end{equation}
with dimensionless positive hyperbolic weights
\begin{equation}
    \label{The_l_i}
    l_i(u) = \frac{\sinh(\Delta_{AD} - u) + b_i \sinh(u)}{\sinh(\Delta_{AD})}.
\end{equation}
Because the null root $b_3 = 0$ leaves the third heavy eigenvalue governed purely by the uncrossed baseline weight, the PD automatically guarantees $l_3(u) > 0$ and $\det[\Lambda_R(u)] \neq 0$. As $\overline{\hat{N}^c_R} = -\hat{N}^T_R \mathcal{C}^{-1}$ transforms with $(\widetilde{U}^\dagger_R)^T = \widetilde{U}^*_R$ under $\hat{N}_R = \widetilde{U}^\dagger_R(\Delta_{AD}) N^{\mathrm{(fl)}}_R$, the flavor-basis Majorana mass matrix and its inverse are given directly by
\begin{equation}
    \nonumber
    \begin{split}
        \mathcal{M}^{\mathrm{(fl)}}_R(u) &= \widetilde{U}^*_R(\Delta_{AD}) \, \Lambda_R(u) \, \widetilde{U}^\dagger_R(\Delta_{AD}), \\
        \left[\mathcal{M}^{\mathrm{(fl)}}_R(u)\right]^{-1} &= \widetilde{U}_R(\Delta_{AD}) \, \Lambda^{-1}_R(u) \, \widetilde{U}^T_R(\Delta_{AD}),
    \end{split}
\end{equation}
with
\begin{equation}
    \nonumber
    \Lambda^{-1}_R(u) = M^{-1}_0 \, \mathrm{diag} \big(s_1 l^{-1}_1 (u), s_2 l^{-1}_2 (u), s_3 l^{-1}_3 (u) \big).
\end{equation}

Integrating out the heavy right-handed eigenstates below the Seesaw threshold via Appelquist--Carazzone decoupling \cite{Appelquist1975} generates the low-energy dimension-five Weinberg operator (see Ref.~\cite{Weinberg1979}), yielding the effective $3 \times 3$ light active neutrino mass matrix in the charged-lepton flavor basis
\begin{equation}
    \label{SMT-I mass matrix}
    \begin{split}
        \mathrm{m}_{\nu} &\approx  -M_D \left[\mathcal{M}_R^{\mathrm{(fl)}}(u)\right]^{-1} M_D^T \\
        &\approx -m^2_D \, \widetilde{U}_R(\Delta_{AD}) \, \Lambda^{-1}_R(u) \, \widetilde{U}^T_R(\Delta_{AD}).
    \end{split}
\end{equation}
Written in the flavor basis aligned with the charged leptons, $\mathrm{m}_\nu$ is Takagi-diagonalized into the positive physical mass basis via the unitary PMNS transformation $U_{\mathrm{PMNS}}$ (see Eq.~(\ref{U_PMNS}) in Sec.~\ref{DPT})
\begin{equation}
    \label{Diag_mass_matrix}
    m_\nu = U^\dagger_{\mathrm{PMNS}} \, \mathrm{m}_{\nu} \, U^*_{\mathrm{PMNS}} = \mathrm{diag}(m_1, m_2, m_3).
\end{equation}

\subsection{Heavy Majorana Mass Eigenvalues}
\label{Der_Mass_Eig}

Within the PD, where the pairwise loop modulus is fixed on the real hyperbolic branch as in Eq.~(\ref{loop_weight}), the affine bond-sum operator projects the static one-magnon roots $b_i$ of Eq.~(\ref{The_b_i}) onto strictly positive hyperbolic trajectories, generating the dimensionless heavy Majorana mass weights $l_i(u)$ in Eq.~(\ref{The_l_i}). Since the heavy states represent physical Majorana fermions, their kinematic masses are given by the absolute magnitudes of the signed eigenvalues in Eq.~(\ref{M_0_eigenvalues})
\begin{equation}
    \label{Majorana_Mass}
    M_i = |\Lambda_i(u)| = M_0 |l_i (u)| = M_0 l_i(u).   
\end{equation}

Using Eqs.~(\ref{M_0_eigenvalues}) and (\ref{Majorana_Mass}) along with Eq. (\ref{loop_weight}), the heaviest Majorana mass simplifies via hyperbolic half-angle identities to
\begin{equation}
    \nonumber
    \begin{split}
        M_1 &= M_0 l_1(u) \\
        &= \frac{M_0}{\sinh(\Delta_{AD})} \big [ \sinh ( \Delta_{AD} - u) + (d'_{AD} + 1) \sinh(u) \big ] \\
        &= M_0 \frac{\sinh\big ( \frac{\Delta_{AD}}{2} + u \big )}{\sinh\big ( \frac{\Delta_{AD}}{2} \big )}.
    \end{split}
\end{equation}
Similarly, the second heavy mass eigenstate is
\begin{equation}
    \nonumber
    \begin{split}
        M_2 &= M_0 l_2(u) \\
        &= \frac{M_0}{\sinh(\Delta_{AD})} \big [ \sinh ( \Delta_{AD} - u) + (d'_{AD} - 1) \sinh(u) \big ] \\
        &= M_0 \frac{\cosh\big ( \frac{\Delta_{AD}}{2} + u \big )}{\cosh\big ( \frac{\Delta_{AD}}{2} \big )},
    \end{split}
\end{equation}
and the third heavy eigenstate is governed purely by the uncrossed identity weight
\begin{equation}
    \nonumber
    M_3 = M_0 l_3(u) = M_0 \frac{\sinh ( \Delta_{AD} - u)}{\sinh(\Delta_{AD})} .
\end{equation}
Given that within the PD
\begin{equation}
    \label{M_!_over_M_2}
    \frac{M_1}{M_2} = \frac{\tanh\big ( \frac{\Delta_{AD}}{2} + u \big )}{\tanh\big ( \frac{\Delta_{AD}}{2} \big ) } > 1,
\end{equation}
and
\begin{equation}
    \nonumber
    \frac{M_2}{M_3} = \frac{\sinh ( \Delta_{AD} ) \cosh\big ( \frac{\Delta_{AD}}{2} + u \big )}{\sinh ( \Delta_{AD} - u ) \cosh\big ( \frac{\Delta_{AD}}{2} \big )} > 1,
\end{equation}
the heavy Majorana mass ordering is strictly locked to
\begin{equation}
     \label{Majorana_Ordering}
     M_1 > M_2 > M_3 > 0.
\end{equation}

The heavy mass-squared splittings are
\begin{equation} \nonumber
    \begin{split}
        \Delta M^2_{21} &= M^2_0 \big(l_2^2 - l_1^2 \big) \\
        &= - 4 M^2_0 \frac{ \sinh(u)\sinh(\Delta_{AD} + u)}{\sinh^2( \Delta_{AD} ) }, \\
        \Delta M^2_{31} &= M^2_0 \big( l_3^2 - l_1^2 \big) \\
        &= -\frac{M_0^2}{\sinh^2(\Delta_{AD})} \Big[ \sinh(2\Delta_{AD})\sinh(2u) \\
        &\quad + 2\sinh(\Delta_{AD} + u)\sinh(u) + \sinh^2(u) \Big],
    \end{split}
\end{equation}
yielding the heavy splitting ratio
\begin{equation} 
    \nonumber
    \frac{\Delta M^2_{21}}{\Delta M^2_{31}} = \frac{4\big[\sinh(\Delta_{AD}) + \cosh(\Delta_{AD})\tanh(u)\big]}{\big[2\cosh(\Delta_{AD}) + 1\big]\big[2\sinh(\Delta_{AD}) + \tanh(u)\big]}.
\end{equation}

\section{RESIDUAL SYMMETRIES, UNITARY BOUNDARY TWIST, AND CP VIOLATION}
\label{DPT}

The heavy Majorana mass operator $\Lambda_R(u)$ in Sec.~\ref{LE_epoch} requires real eigenvalues. Thus, the Temperley--Lieb generators $e_i$ of Eq.~(\ref{B_operator_sum}) are forced to be Hermitian, $e^\dagger_i = e_i$. Because the Temperley--Lieb loop modulus $d = q + q^{-1}$ generates real eigenvalues $d \pm 1$ and $0$ for any real $d$, the requirement of real Majorana masses allows alternative real branches, such as the trigonometric $d = 2\cos(\Delta_{AD})$ or the icosahedral $d = \phi$. However, to smoothly and analytically connect the topological defect $\Delta_{AD}$ to the unbraided, scale-invariant asymptotic limit $\Delta_0 \to 0$, with $d_0 = 2$, the architecture structurally selects the hyperbolic branch $q'_{AD} = e^{\Delta_{AD}}$, yielding the loop weight in Eq. (\ref{loop_weight}). Hence, a basis redefinition of the fermion fields, $\hat{N}_R \mapsto U \hat{N}_R$, must be unitary, $U^\dagger U = I$, to preserve the canonical normalization of the fermion kinetic terms.

The value of the elementary angular defect $\Delta_{AD}$ is established via the first postulate (P1) of the architecture: Weyl-averaged chamber deficit, by linking the discrete curvature of an icosahedral vertex to the six-chamber Coxeter complex of the three-generation algebra $\mathcal{H}_3(q'_{AD})$ (see Ref.~\cite{Jones1987} and Ch.~7 in Ref.~\cite{Humphreys1990}). While the curved icosahedral surface has five equilateral triangles of internal angle $\pi/3$ meeting at each $\mathbb{Z}_5$-stabilized vertex, its discrete Gauss--Bonnet curvature $\Omega_{\mathrm{vertex}} = 2\pi - 5(\pi/3) = \pi/3$ (see Refs.~\cite{Banchoff1967, Coxeter1973}) is defined relative to the flat two-dimensional Euclidean tangent plane $T_v \cong \mathbb{R}^2$, which is tiled by the six equilateral-triangle Weyl chambers of the planar $A_2$ Coxeter complex: $6 \times \pi/3 = 2\pi$. Geometrically, folding the flat $A_2$ tangent star into the three-dimensional icosahedral vertex cone excises one of the six $\pi/3$ chambers. Since the three-generation operator algebra $\mathcal{H}_3(q'_{AD})$ lives on the $A_2$ Coxeter complex: $\dim \mathcal{H}_3(q'_{AD}) = |W(A_2)| = |S_3| = 6$, rather than on $\mathbb{Z}_5$, pulling the vertex deficit back to the flat tangent star and averaging over the simply transitive action of $W(A_2) = S_3$ across all six tangent chambers, including the excised sector, yields the Weyl-averaged chamber deficit
\begin{equation}
    \label{The_D_AD}
    \Delta_{AD} = \frac{\Omega_{\mathrm{vertex}}}{|W(A_2)|} = \frac{\pi/3}{3!} = \frac{\pi}{18}.
\end{equation}

The unitary boundary twist $U_{\mathrm{twist}}$ that rotates the heavy-state basis relative to $U_{\mathrm{GR}}$ is not an element of the Temperley--Lieb representation of $\mathcal{H}_3(q'_{AD})$. In the three-dimensional representation $\hat{\mathcal{H}}$ generated by $S_i = q'_{AD}I - e_i$, the one-dimensional $q$-antisymmetrizer ideal $A = \sum_{w \in S_3} (-q'_{AD})^{-\ell(w)} S_w$ vanishes identically: $A = 0$, by the standard Iwahori--Hecke to Temperley--Lieb quotient (see Ref.~\cite{Jones1987} and Sec.~10.2 in Ref.~\cite{Chari1994}). Consequently, by Schur's lemma and the representation theory of Temperley--Lieb algebras \cite{Goodman1989}, every operator in the image of $\mathcal{H}_3(q'_{AD})$ is $2 \oplus 1$ block-diagonal in the $B$-eigenbasis, mixing only $|b_1\rangle$ and $|b_2\rangle$ within the loop module $\mathcal{V}_2$ while leaving the singlet $|b_3\rangle \in \mathcal{V}_1$ unmixed. Instead, the flavor-basis transformation $\widetilde{U}_R(\Delta_{AD})$ is obtained directly from the residual symmetries of the $A_5 \rtimes$ CP flavor manifold following the formalism of \cite{Feruglio2013, Ding2012}.

Therefore, by the second postulate (P2): residual symmetries, twist angle, and Majorana parities, the charged-lepton mass basis preserves the cyclic subgroup $\mathbb{Z}_5 = \langle \mathcal{T} \rangle$ with $\mathcal{T} = \mathrm{diag}(1, q^2_H, q^{-2}_H)$, while the $\sigma$-symmetric Majorana neutrino parent preserves $\mathbb{Z}_2 \times \mathrm{CP} = \langle G_2 \rangle \times \langle X \rangle$ with $(1,3)$ rotation magnitude $|\theta| = \Delta_{AD}$. In the $V_4$ eigenbasis defined by the columns of $U_{\mathrm{GR}}$, the preserved $\mathbb{Z}_2$ generator is $\hat{G}_2 = \mathrm{diag}(-1, +1, -1)$, such that $G_2 = U_{\mathrm{GR}} \hat{G}_2 U^T_{\mathrm{GR}} \in A_5$ fixes the second column of $U_{\mathrm{GR}}$; and the generalized $\mu$--$\tau$ reflection~\cite{Harrison2002_2, Grimus2004, Xing2016} CP transformation $X$ is determined by the third column $c_3 = (0, 1/\sqrt{2}, 1/\sqrt{2})^T$ of $U_{\mathrm{GR}}$
\begin{equation}
    \nonumber
    X = I - 2 c_3 c^T_3 = \begin{pmatrix} 1 & 0 & 0 \\ 0 & 0 & -1 \\ 0 & -1 & 0 \end{pmatrix} = \Sigma \Sigma^T,
\end{equation}
with $\Sigma = U_{\mathrm{GR}} \, \mathrm{diag}(1, 1, -i)$. 

Bond exchange $\sigma$ acts with opposite signs on the loop-coupled pair rather than mixing them; off-diagonal mixing between these states is instead induced by the bond-asymmetric operator $e_1 - e_2$. Furthermore, identifying CP eigenvalues with the bond-exchange action fixes opposite signs for the loop-coupled solar pair: $s_1 = +1$, $s_2 = -1$, while leaving the decoupled singlet sign unfixed by exchange alone; P2 declares $s_3 = -1$, fixing the discrete CP eigenvalues to
\begin{equation} \label{d_AD_prime}
    s_i = (+1, -1, -1).
\end{equation}

By the general theorem of Ref.~\cite{Feruglio2013}, generalized CP invariance of the light neutrino mass matrix under
\begin{equation}
    \label{Gen_CP_Inv}
     G^T_2 \mathrm{m}_\nu G_2 = \mathrm{m}_\nu, \text{ and } X \mathrm{m}^*_\nu X = \mathrm{m}_\nu,
\end{equation}
restricts the basis transformation $\widetilde{U}_R$ satisfying the canonical CP condition $X \widetilde{U}^*_R = \widetilde{U}_R$ to a one-parameter family of rotations
\begin{equation} \nonumber
    R_{13}(\theta) = \begin{pmatrix} \cos\theta & 0 & -\sin\theta \\ 0 & 1 & 0 \\ \sin\theta & 0 & \cos\theta \end{pmatrix},
\end{equation}
in the $-1$ eigenplane of $\hat{G}_2$, spanned by heavy states $1$ and $3$,
\begin{equation}
    \label{FHZ_form}
    \widetilde{U}_R(\theta) = \Sigma \, R_{13}(\theta),
\end{equation}
such that the observable mixing matrix takes the form $U_{\mathrm{PMNS}} = \Sigma R_{13}(\theta) P$, where $P$ is a diagonal matrix of discrete CP-allowed Majorana phase factors. The following resulting predictions are direct consequences of this parametrization: the CP-conserving Majorana phases (see Sec. \ref{AIMP}); and $\theta_{23}$, $|\delta_{\mathrm{CP}}|$, and the solar-reactor correlation (see Sec. \ref{PMNSMA}).

Because in the heavy-state label basis the Hermitian involution with $\det = +1$: $S^{(0)}_{\max} = -p_{13}$ (see App.~\ref{TUMCI} for involution details), commutes with $\hat{G}_2$: $[S^{(0)}_{\max}, \hat{G}_2] = 0$, and generates rotations in the $(1,3)$ eigenplane, Eq.~(\ref{FHZ_form}) connects to the unitary boundary twist operator $U_{\mathrm{twist}}$. Under P2, the magnitude of the $(1,3)$ rotation angle is identified with the elementary angular defect $|\theta| = \Delta_{AD}$. Resolving $U_{\mathrm{twist}}$ in $\mathrm{span}\{I, S^{(0)}_{\max}\}$ with trigonometric weights $\cos(\Delta_{AD})$ and $\sin(\Delta_{AD})$ and relative phase $e^{\pm i\delta_{\mathrm{CP}}}$ (see Refs.~\cite{Kitaev2003, Kauffman1987})
\begin{equation} 
    \nonumber
    U_{\mathrm{twist}} = \cos(\Delta_{AD}) I + \sin(\Delta_{AD}) e^{\pm i \delta_{\mathrm{CP}}} S^{(0)}_{\max},
\end{equation}
and enforcing unitarity across the basis transformation
\begin{equation}
    \nonumber
    U_{\mathrm{twist}} U^{\dagger}_{\mathrm{twist}} = I + \sin(2\Delta_{AD})\cos(\delta_{\mathrm{CP}}) S^{(0)}_{\max} = I,
\end{equation}
yields
\begin{equation}
    \label{This_is_delta_CP}
     \cos(\delta_{\mathrm{CP}}) = 0, \qquad \delta_{\mathrm{CP}} = \pm \pi/2.
\end{equation}

By $(S^{(0)}_{\max})^2 = I$, this unitary operator resums identically into Stone's exponential map for the two chiral branches $\theta = \pm\Delta_{AD}$
\begin{equation}
    \nonumber
    \begin{split}
        U_{\mathrm{twist}}(\pm\Delta_{AD}) &= e^{\pm i \Delta_{AD} S^{(0)}_{\max}} \\
        & = \cos(\Delta_{AD}) I \pm i \sin(\Delta_{AD}) S^{(0)}_{\max} \\
        &= \begin{pmatrix}
        c_{AD} & 0 & \mp i s_{AD} \\
        0 & c_{AD} \mp i s_{AD} & 0 \\
        \mp i s_{AD} & 0 & c_{AD}
    \end{pmatrix},
    \end{split}
\end{equation}
where $c_{AD} = \cos(\Delta_{AD})$ and $s_{AD} = \sin(\Delta_{AD})$. Comparing this direct expansion to the standard phenomenological CP phase parametrization identifies the phase factor as $e^{i\delta_{\mathrm{CP}}} = \pm i$, immediately forcing $\delta_{\mathrm{CP}} = \pm \pi/2$. Because $S^{(0)}_{\max}$ has eigenvalues $\pm 1$, $U_{\mathrm{twist}}$ has eigenvalues $q^{\pm 1}_{AD} = e^{\pm i\Delta_{AD}}$ and satisfies the unitary rotation minimal polynomial
\begin{equation} \nonumber
    \begin{split}
        (U_{\mathrm{twist}} - q_{AD}I)(U_{\mathrm{twist}} - q^{-1}_{AD}I) &= U^2_{\mathrm{twist}} - d_{AD} U_{\mathrm{twist}} + I \\
        &= 0,
    \end{split}
\end{equation}
where
\begin{equation}
    \label{2cos_Delta}
    d_{AD} = q_{AD} + q^{-1}_{AD} = 2\cos(\Delta_{AD}).
\end{equation}

Right-multiplying $U_{\mathrm{twist}}$ on $U_{\mathrm{GR}}$ defines the un-rephased matrix $U_R(\Delta_{AD}) = U_{\mathrm{GR}} U_{\mathrm{twist}}$
\begin{widetext}
\begin{equation}
    \nonumber
    U_R(\Delta_{AD}) = \begin{pmatrix}
        c_{AD} \, \phi_5 & (c_{AD} - s_{AD} e^{\pm i \delta_{\mathrm{CP}}}) \phi_1 & - s_{AD} e^{\pm i \delta_{\mathrm{CP}}} \phi_5 \\
        - c_{AD} \, \phi_2 - \frac{s_{AD} e^{\pm i \delta_{\mathrm{CP}}}}{\sqrt{2}} & (c_{AD} - s_{AD} e^{\pm i \delta_{\mathrm{CP}}}) \frac{\phi_5}{\sqrt{2}} &  s_{AD} e^{\pm i \delta_{\mathrm{CP}}} \phi_2 + \frac{c_{AD}}{\sqrt{2}} \\
        c_{AD} \, \phi_2 - \frac{s_{AD} e^{\pm i \delta_{\mathrm{CP}}}}{\sqrt{2}} & -(c_{AD} - s_{AD} e^{\pm i \delta_{\mathrm{CP}}}) \frac{\phi_5}{\sqrt{2}} & - s_{AD} e^{\pm i \delta_{\mathrm{CP}}} \phi_2 + \frac{c_{AD}}{\sqrt{2}}
    \end{pmatrix},
\end{equation}
\end{widetext}
with GR coefficients
\begin{equation} \nonumber
    \phi_1 = \sqrt{\frac{1}{\sqrt{5} \phi}}, \qquad
    \phi_2 = \sqrt{\frac{1}{2\sqrt{5} \phi}}, \qquad
    \phi_5 = \sqrt{\frac{\phi}{\sqrt{5}}}.
\end{equation}
Given that
\begin{equation} \nonumber
    \begin{split}
        U^\dagger_R(\Delta_{AD}) X U^*_R(\Delta_{AD}) &= \mathrm{diag}(1, e^{\pm 2i\Delta_{AD}}, -1) \\
        &= K^2_e(\pm\Delta_{AD}), \\
        \\
        K_e(\pm\Delta_{AD}) &= \mathrm{diag}(1, e^{\pm i\Delta_{AD}}, \pm i),
    \end{split}
\end{equation}
enforcing the canonical $\mathbb{Z}_2 \times$ CP invariance condition
\begin{equation}
    \label{Inv_Con}
    X \widetilde{U}^*_R(\Delta_{AD}) = \widetilde{U}_R(\Delta_{AD}),
\end{equation}
uniquely fixes
\begin{equation}
    \label{U_R_Tilde}
    \widetilde{U}^{(\pm)}_R(\Delta_{AD}) = U_{\mathrm{GR}} U_{\mathrm{twist}}(\pm\Delta_{AD}) K_e(\pm\Delta_{AD}),
\end{equation}
which satisfies the algebraic identity
\begin{equation}
    \label{Equiv_Twist_FHZ}
    \begin{split}
        U_{\mathrm{twist}}&(\pm\Delta_{AD}) \, K_e(\pm\Delta_{AD}) \\
        &= \mathrm{diag}(1, 1, -i) \, R_{13}(\pm\Delta_{AD}) \, \mathrm{diag}(1, 1, \mp 1).
    \end{split}
\end{equation}
While exponentiating $S_{\max}$ rigorously yields the appropriate sum of the sine and cosine infinite series to construct the unitary rotation, the specific identification of this rotation angle with the geometric defect, $|\theta| = \Delta_{AD}$, is adopted via P2.

Eq. (\ref{U_R_Tilde}), by means of Eq.~(\ref{Equiv_Twist_FHZ}), is identical to the canonical $\mathbb{Z}_2 \times$ CP matrix of Eq.~(\ref{FHZ_form}) up to the real sign $\mp 1$ on the third column, with a strictly real and positive electron row $\widetilde{U}_{R, ei} = (c_{AD}\phi_5, \, \phi_1, \, s_{AD}\phi_5)$. Including the diagonal Majorana phase matrix $P$ yields the complete PMNS mixing matrix
\begin{equation}
    \label{U_PMNS}
    U_{\mathrm{PMNS}} = \widetilde{U}_R(\Delta_{AD}) P = U_{\mathrm{GR}} U_{\mathrm{twist}} K_e P.
\end{equation}

\subsection{Discrete CP Eigenvalues and Majorana Phases}
\label{AIMP}

In the canonical $\mathbb{Z}_2 \times$ CP basis $\widetilde{U}_R(\Delta_{AD})$ where Eq.~(\ref{Inv_Con}) holds, generalized CP invariance, Eq. (\ref{Gen_CP_Inv}), requires the signed heavy eigenvalues $\Lambda_i(u) = s_i M_i(u)$ to carry discrete CP eigenvalues $s_i$ (see Refs.~\cite{Wolfenstein1981, Feruglio2013})
\begin{equation}
    \nonumber
    e^{i\varphi_i} = \frac{\Lambda_i(u)}{M_i(u)} = \frac{\Lambda_i(u)}{|\Lambda_i(u)|} = s_i.
\end{equation}
Under P2, via Eq. (\ref{d_AD_prime}), the three polar orientation factors evaluate directly to
\begin{equation} \nonumber
    \begin{split}
        \frac{\Lambda_1}{M_1} &= s_1 = +1 = e^{i\varphi_1}, \\
        \frac{\Lambda_2}{M_2} &= s_2 = -1 = e^{i\varphi_2}, \\
        \frac{\Lambda_3}{M_3} &= s_3 = -1 = e^{i\varphi_3}. 
    \end{split}
\end{equation}
Thus, the internal Majorana phases $\varphi_i$ are quantized to the CP-conserving values $\varphi_1 = 0$, $\varphi_2 = \pi$, and $\varphi_3 = \pi$. In the Takagi diagonalization of $\mathrm{m}_\nu$ in Eq.~(\ref{Diag_mass_matrix}), enforcing positive physical light masses $m_i > 0$ absorbs the square roots of the Seesaw-inverted signs $-s_i = (-1, +1, +1)$ into the diagonal Majorana phase matrix. Selecting the unlooped null root $b_3 = 0$ as the phase reference and defining the relative Majorana phases via the invariant ratios
\begin{equation}
    \label{Rel_phases}
    e^{i(\varphi_i - \varphi_j)} = e^{i \alpha_{ij}} = \frac{M_j \Lambda_i}{M_i \Lambda_j} = \frac{s_i}{s_j},
\end{equation}
fixes the observable relative Majorana phases with respect to $\varphi_3 = \pi$ to
\begin{equation}
    \label{The_alphas}
    \begin{split}
        \alpha_{13} &= \varphi_1 - \varphi_3 = -\pi \equiv \pi \pmod{2\pi}, \\
        \alpha_{23} &= \varphi_2 - \varphi_3 = 0.
    \end{split}
\end{equation}
Consequently, the relative Majorana phase matrix in Eq.~(\ref{U_PMNS}) takes the form
\begin{equation} \nonumber
    P = \mathrm{diag}(e^{i\alpha_{13}/2}, e^{i\alpha_{23}/2}, 1) = \mathrm{diag}(-i, 1, 1),
\end{equation}
completing the full observable $U_{\mathrm{PMNS}}$. The PDG parametrization (see Sec.~14 in Ref.~\cite{PDG2024}): $P_{\mathrm{PDG}} = \mathrm{diag}(e^{(i \eta_1)}, e^{(i \eta_2)}, 1)$, has $\eta_{1} = 0$ and $\eta_{2} = \pi/2$ (mod 2$\pi$).

\section{ANALYTICAL PMNS MIXING ANGLES}
\label{PMNSMA}

The mixing angles in the standard PDG parameterization $U^{s}_{\mathrm{PMNS}}$ follow directly from the matrix elements of $U_{\mathrm{PMNS}}$ in Eq.~(\ref{U_PMNS}) evaluated at $\Delta_{AD} = \pi/18$. Equating $|U^s_{e3}|^2$ to $|U_{e3}|^2$ determines the reactor angle (following the convention $c_{ij} = \cos{(\theta_{ij})}$ and $s_{ij} = \sin{(\theta_{ij})}$)
\begin{equation}
    \label{Theta_13}
    \begin{split}
        \big| s_{13} e^{\mp i \delta_{\mathrm{CP}}} \big|^2 &= \left| s_{AD} \sqrt{\frac{\phi}{\sqrt{5}}} \right|^2, \\
        s_{13} &= s_{AD} \sqrt{\frac{\phi}{\sqrt{5}}}, \\
        \theta_{13} &= 8.49^\circ.
    \end{split}
\end{equation}
Without an explicit UV completion, the primary empirical test of P1 is $\theta_{13}$: whereas Eq.~(\ref{The_D_AD}) predicts $\sin^2{(\theta_{13})} = 0.0218$, which sits at $-1.1\sigma$ of the NuFit 6.1 central value $0.0225 \pm 0.0006$ (see Refs.~\cite{nufit2024, NuFit61}), the adjacent unit fractions $\pi/17$, with $\sin^2{(\theta_{13})} = 0.0244$, and $\pi/19$, with $\sin^2{(\theta_{13})} = 0.0196$, are in tension with the latest global fits (see Refs.~\cite{nufit2024, NuFit61}) at the $+3.2\sigma$ and $-4.8\sigma$ levels, respectively.

Consider the five Platonic solids with their proper rotational symmetry groups in $SO(3)$ and vertex angular deficits (see Refs. ~\cite{Coxeter1973, Banchoff1967}): the tetrahedron ($A_4$, $\Omega = \pi$), cube ($S_4$, $\Omega = \pi/2$), octahedron ($S_4$, $\Omega = 2\pi/3$), dodecahedron ($A_5$, $\Omega = \pi/5$), and icosahedron ($A_5$, $\Omega = \pi/3$), paired with the standard one-angle $\mathbb{Z}_2 \times$ CP lepton mixing patterns of these discrete groups (see Refs.~\cite{Altarelli2010, King2013, Feruglio2013, Ding2026, Petcov2026}), and integer divisors $n \in \{1, \dots, 6\}$. Across this broad set of 62 recipes of which 39 yield $\theta_{13} \le 20^\circ$, exactly two constructions land inside $[8.27^\circ, 8.95^\circ]$: the present icosahedral $A_5$ prescription and the cubic $S_4$ $\mathrm{TM}_1$ prescription, with: $\theta = (\pi/2)/6 = \pi/12$, which yields: $\theta_{13} = 8.59^\circ \implies \sin^2{(\theta_{12})} = 0.3181$. Under a naive uniform distribution on $[0^\circ, 20^\circ]$, the probability of at least one hit across $\sim 39$ candidates is $\sim 0.74$. If the divisor is instead fixed \textit{a priori} to $n = |W(A_2)| = 6$, where 10 combinations yield $\theta_{13} \le 20^\circ$, the same two hits remain. Distinguishing the icosahedral construction from the cubic $S_4$ competitor therefore rests on the geometric requirement in P1. This restricts the set to the tetrahedron, octahedron, and icosahedron, where the $S_4$ octahedron gives $\theta = \pi/9$ and $\sin^2\theta_{13} = 0.0390$. Accordingly, the agreement of P1 with $\theta_{13}$ is not by itself a statistical proof against coincidence. It serves as a falsifiable geometric ansatz tested with the mixing correlations presented in this Section.

The solar angle is determined by equating $|U^s_{e2}/U^s_{e1}|^2$ to $|U_{e2}/U_{e1}|^2$ from Eq.~(\ref{U_PMNS})
\begin{equation}
    \label{Theta_12}
    \begin{split}
        \left| \frac{s_{12} c_{13}}{c_{12} c_{13}} \right|^2 &= \left| \frac{\sqrt{\frac{1}{\sqrt{5} \phi}}}{c_{AD} \sqrt{\frac{\phi}{\sqrt{5}}}} \right|^2, \\
        \tan(\theta_{12}) &= \frac{ 1 }{\phi \, c_{AD}}, \\
        \theta^{(s)}_{12} &= 32.11^\circ.
    \end{split}    
\end{equation}
Since the $(1,3)$ rotation leaves the modulus of the second column of $U_{\mathrm{GR}}$ invariant, the architecture enforces the exact solar-reactor correlation
\begin{equation} \label{Solar_cor}
    \sin^2{(\theta_{12})}\cos^2{(\theta_{13})} = (\sqrt{5}\phi)^{-1},
\end{equation}
lifting $\theta_{12}$ by $+0.39^\circ$ above the unperturbed GR baseline $\theta^{(0)}_{12} = 31.72^\circ$ (see Refs.~\cite{Everett2009, Ding2012}). However, the symmetric baseline value from Eq. (\ref{Theta_12}), corresponding to $\sin^2{(\theta^{(s)}_{12})} = 0.2826$, is now disfavored by the latest JUNO measurement~\cite{JUNO2026}: $(\sin^2{(\theta_{12})})_{\text{JUNO}} = 0.3092 \pm 0.0087$ (at $1\sigma$), by more than $3\sigma$. This tension is confronted directly in Sec. \ref{App_Extension} by introducing an asymmetric-bond extension that shifts the prediction into agreement with current data.

The atmospheric angle is determined by equating $|U^s_{\mu 3}/U^s_{\tau 3}|^2$ to $|U_{\mu 3}/U_{\tau 3}|^2$ from Eq.~(\ref{U_PMNS})
\begin{equation}
    \label{Theta_23}
    \begin{split}
        \left| \frac{s_{23} c_{13}}{c_{23} c_{13}} \right|^2 &= \left| \frac{ \pm i s_{AD} \phi_2 + \frac{c_{AD}}{\sqrt{2}}}{\mp i s_{AD} \phi_2 + \frac{c_{AD}}{\sqrt{2}}} \right|^2 = 1, \\
        \tan(\theta_{23}) &= 1, \\
        \theta_{23} &= 45^\circ,
    \end{split}
\end{equation}
yielding exact maximal atmospheric mixing.

The unitarity and generalized CP condition $\cos(\delta_{\mathrm{CP}}) = 0$ in Eq.~(\ref{This_is_delta_CP}) fixes the magnitude $|\delta_{\mathrm{CP}}| = \pi/2$ while leaving an exact twofold chiral sign degeneracy, $\delta_{\mathrm{CP}} = \pm \pi/2$, corresponding to the two rotation signs $\theta = \pm\Delta_{AD}$ in Eq.~(\ref{Equiv_Twist_FHZ}).

\subsection{Asymmetric Bond Extension and the Solar Angle}
\label{App_Extension}

The foundational baseline of this architecture restricts the heavy Majorana mass operator to the symmetric affine subspace $\mathrm{span}\{I, \, e_1 + e_2\}$, which is invariant under the bond-exchange automorphism  $\sigma: e_1 \leftrightarrow e_2$ and algebraically fixes the solar mixing angle in Eq.~(\ref{Theta_12}). However, the current tension with the latest JUNO measurements in \cite{JUNO2026} disfavors the symmetric result in Eq. (\ref{Theta_12}) by more than $3\sigma$. To account for this disagreement with experiment, the five-dimensional Temperley--Lieb algebra $TL_3(d'_{AD})$ (see Ref.~\cite{Goodman1989}) provides exactly one unused real-symmetric basis direction orthogonal to $\mathrm{span}\{I, \, B\}$: the antisymmetric operator $e_1 - e_2$.

Relaxing exact $\sigma$-symmetry by allowing the two nearest-neighbor bonds to carry unequal weights yields an algebraic extension of the mass operator (see Ref.~\cite{Goodman1989})
\begin{equation}
    \label{New_R}
    R = a I + b(e_1 + e_2) + c(e_1 - e_2).
\end{equation}
In the $B$-eigenbasis, $e_1 - e_2$ is strictly off-diagonal, mixing only the loop-coupled eigenstates $|b_1\rangle$ and $|b_2\rangle$ while annihilating the trivial singlet $|b_3\rangle$. Diagonalizing the extended operator therefore introduces a real orthogonal rotation $R_{12}(\varepsilon)$ acting in the $(1,2)$ heavy mass eigenbasis. While $TL_3(d'_{AD})$ isolates the $(1,2)$ rotation plane, it does not fix the continuous angle $\varepsilon$. To eliminate this free parameter without introducing an arbitrary continuous fit, the fourth postulate (P4): asymmetric-bond solar rotation, fixes the rotation angle at quadratic order in the primary topological defect
\begin{equation}
    \label{perturbation}
    \varepsilon = \Delta_{AD}^2 = (\pi/18)^2 \approx 0.03046 \text{ rad}.
\end{equation}

Eq.~(\ref{perturbation}) is not an algebraic consequence of an $SO(3)$ group commutator: the baseline architecture contains only a single $\mathcal{O}(\Delta_{AD})$ rotation, $R_{13}(\Delta_{AD})$. Moreover, if an auxiliary $\mathcal{O}(\Delta_{AD})$ rotation in the $(2,3)$ plane were supplied, the resulting group commutator would carry $\mathcal{O}(\Delta_{AD}^3)$ corrections that shift $\sin^2{(\theta_{13})}$ away from $0.0218$ to $0.0208$ or $0.0229$, depending on sign convention. P4 activates the unique orthogonal real-symmetric generator $e_1 - e_2$ of $TL_3(d'_{AD})$ via a right-multiplied $(1,2)$ rotation $\widetilde{U}_R \to \widetilde{U}_R R_{12}(\varepsilon)$ with the fixed angle in Eq. (\ref{perturbation}). This preserves the generalized $\mu$--$\tau$ reflection CP condition $X (\widetilde{U}_R R_{12})^* = \widetilde{U}_R R_{12}$ exactly while softly breaking the residual $\mathbb{Z}_2$ symmetry $\hat{G}_2$. In Eq. (\ref{New_R}) $a$ and $b$ carry the baseline weights of the symmetric operator in Eq.~(\ref{Final_R}) on the PD, while $c$ parametrizes the bond-asymmetric departure. Postulate P4 does not fit $c$ numerically, it requires Eq. (\ref{perturbation}).

Consequently, $\theta_{23} = 45^\circ$, $|\delta_{\mathrm{CP}}| = \pi/2$, and the reactor element $U_{e3} = s_{AD}\phi_5$ remain strictly invariant, while the first two columns of the electron row mix linearly
\begin{equation}
    \nonumber
    \begin{split}
        U'_{e2} &= U_{e1}\sin{(\Delta_{AD}^2)} + U_{e2}\cos{(\Delta_{AD}}^2) \\
        &= c_{AD}\phi_5\sin{(\Delta_{AD}^2)} + \phi_1\cos{(\Delta_{AD}^2)}.
    \end{split}
\end{equation}
Thus, the solar angle becomes
\begin{equation}
    \label{Extended_Solar}
    \begin{split}
        \sin^2{(\theta'_{12})} &= \frac{|U'_{e2}|^2}{\cos^2{(\theta_{13})}} \\
        &= \frac{\left[c_{AD}\phi_5\sin{(\Delta_{AD}^2)} + \phi_1\cos{(\Delta_{AD}^2)}\right]^2}{1 - s^2_{AD}\phi^2_5}.
    \end{split}
\end{equation}
Evaluating Eq.~(\ref{Extended_Solar}) at $\varepsilon = \Delta_{AD}^2$ shifts the solar prediction from the symmetric baseline $0.2826$ to
\begin{equation}
    \label{The_real_Theta_12}
    \sin^2{(\theta'_{12})} = 0.3104,
\end{equation}
in agreement with the latest JUNO result $(\sin^2{(\theta_{12})})_{\text{JUNO}} = 0.3092 \pm 0.0087$ in \cite{JUNO2026}, which already places strong statistical pressure on unperturbed $A_5$ mixing patterns~\cite{Petcov2026}. This result is used in the 6.1 NuFit \cite{nufit2024, NuFit61}: $(\sin^2{(\theta_{12})})_{\mathrm{NuFit}} = 0.3088^{+0.0207}_{-0.0195}$ (at $3\sigma$), and is used in \cite{Capozzi2026, Ding2026, He2026} to test its implications. The agreement between the value in Eq. (\ref{The_real_Theta_12}) with the latest JUNO data carries no \textit{a priori} statistical weight, and its genuine empirical test lies in JUNO's forthcoming 6 year-exposure precision [0.302, 0.312], at $3\sigma$ \cite{JUNO2022}.

In contrast to the $\sigma$-symmetric $A_5$ relation in Eq. (\ref{Solar_cor}), standard $\mathbb{Z}_2 \times$ CP trimaximal constructions in $A_4$ and $S_4$ point in distinct directions. A $\mathrm{TM}_2$ alignment, which preserved the second column, common to $A_4$ and $S_4$, enforces $\sin^2\theta_{12}\cos^2\theta_{13} = 1/3$, giving $\sin^2\theta_{12} \approx 0.3408$ at $\sin^2\theta_{13} = 0.0218$ and thus overshooting JUNO's central value. Conversely, a $\mathrm{TM}_1$ alignment, which preserves the first column, characteristic of $S_4$, gives $\sin^2\theta_{12} = 1 - \frac{2}{3\cos^2\theta_{13}} \approx 0.3185$ at $\sin^2\theta_{13} = 0.0218$ (or $\approx 0.3181$ at $\theta = \pi/12$), lying within current JUNO uncertainties without sharing the equilateral-triangular icosahedral face matching of P1. Forthcoming precision measurements from JUNO will directly test the committed $A_5$ prediction of Eq. (\ref{The_real_Theta_12}) against competing symmetry constructions.

Comparing simple ans\"atze built from the architecture's existing parameters with $\varepsilon = \Delta_{AD}/6$ ($\sin^2{(\theta'_{12})} = 0.3091$), $\varepsilon = \Delta_{AD} - u$ ($\sin^2{(\theta'_{12})} = 0.3133$), and $\varepsilon = \sin^2{(\theta_{13})}$ ($\sin^2{(\theta'_{12})} = 0.3024$), against JUNO at its 6 year-exposure precision, shows that these alternatives lie at roughly $-0.8\sigma$, $+1.9\sigma$, and $-5.3\sigma$ from the P4 value, respectively. Hence, 6 years of JUNO data can decisively exclude alternatives such as $\varepsilon = \sin^2{(\theta_{13})}$ and test $\Delta_{AD} - u$. Although JUNO's 6 year-exposure sensitivity will not cleanly separate $\varepsilon = \Delta_{AD}^2$ from $\varepsilon = \Delta_{AD}/6$, P4 adopts $\Delta_{AD}^2$ on grounds of perturbative naturalness: as $\Delta_{AD}$ is a small dimensionless defect parameter, a quadratic correction of order $\Delta_{AD}^2$ follows a systematic expansion in the defect itself without introducing an ad hoc numerical divisor with no geometric origin in the manifold's construction such as $1/6$.

With the PMNS mixing parameters presented in Eqs. (\ref{This_is_delta_CP}, \ref{Theta_13}, \ref{Theta_23} and \ref{The_real_Theta_12}), the basis-independent Jarlskog invariant \cite{Jarlskog2005} is fixed in magnitude
\begin{equation}
    \nonumber
    \begin{split}
        J_{\mathrm{CP}} &= c_{12} s_{12} c_{23} s_{23} c^2_{13} s_{13} \sin(\delta_{\mathrm{CP}}) \\
        &= J^{\max}_{\mathrm{CP}} \sin(\delta_{\mathrm{CP}}) = \pm 3.34 \times 10^{-2}, 
    \end{split}
\end{equation}
as the architecture does not select \textit{a priori} the discrete sign. Thus, determining which chiral orientation is realized in nature is strictly the domain of oscillation experiments. While T2K data (see Ref.~\cite{T2K2025}) favor the negative branch near $\delta_{\mathrm{CP}} \approx -\pi/2$, where $-\pi /2\equiv 3\pi/2$, NOvA data (see Ref.~\cite{NOvA2026}) exhibit tension with this value under the NH, and joint analyses (see Ref.~\cite{NOvAT2K2025}) remain limited by degeneracies between $\delta_{\mathrm{CP}}$, the $\theta_{23}$ octant, and the mass ordering. The maximal atmospheric mixing, the NH, and the maximal CP violation of the architecture provide a sharp two-branch CP-target for next-generation neutrino oscillation experiments (see Refs.~\cite{DUNE2020, Hyper-Kamiokande2018}).

\subsection{Astrophysical Neutrino Telescopes}

For both chiral branches, the residual $\mu$--$\tau$ reflection symmetry $X$ enforces exact modulus symmetry, $|U_{\mu i}|^2 = |U_{\tau i}|^2 = (1 - |U_{ei}|^2)/2$ (see Ref.~\cite{Feruglio2013}). In high-energy astrophysical neutrino telescopes, where oscillation phases average out over cosmological baselines, the flavor-transition probability matrix $P_{\alpha\beta} = \sum_{i=1}^3 |U_{\alpha i}|^2 |U_{\beta i}|^2$ is completely determined by a single parameter, $P_{ee}$ \cite{Pakvasa2008, Bustamante2015}
\begin{equation}
    \nonumber
    \begin{split}
        P_{ee} &= \sum_{i=1}^3 |U_{ei}|^4 = 0.5477, \\
        P_{e\mu} &= P_{e\tau} = \frac{1 - P_{ee}}{2} = 0.2261, \\
        P_{\mu\mu} &= P_{\mu\tau} = P_{\tau\tau} = \frac{1 + P_{ee}}{4} = 0.3869.
    \end{split}
\end{equation}
Convolving $P_{\alpha\beta}$ with the three standard astrophysical production benchmarks in~\cite{Pakvasa2008, Bustamante2015} yields sign-independent earthbound flavor ratios $(f_e : f_\mu : f_\tau)_\oplus$ of $1 : 1 : 1$ (exact) for a canonical pion-decay source, $(1 : 2 : 0)_{\mathrm{source}}$; $0.23 : 0.39 : 0.39$ for a muon-damped source, $(0 : 1 : 0)_{\mathrm{source}}$; and $0.55 : 0.23 : 0.23$ for a neutron-beam source, $(1 : 0 : 0)_{\mathrm{source}}$, offering a clean consistency test for IceCube-Gen2~\cite{IceCubeGen2_2020} and KM3NeT~\cite{KM3Net2016}.

As demonstrated explicitly in App.~\ref{RGE_I}, exact $\mathbb{Z}_2 \times$ CP invariance in the neutrino sector protects the mixing structure across the heavy partial-decoupling thresholds, while the opposite CP eigenvalues of the solar pair, $s_1 s_2 = -1$, and the orthogonality between the Majorana and Dirac phases suppress the low-energy charged-lepton one-loop renormalization group (RG) beta functions to $\dot{\theta}_{ij} \sim 10^{-7}\text{ rad}$ per unit $\ln\mu$.

\section{ACTIVE NEUTRINO MASS SPECTRUM}
\label{AFAM}

Having established the heavy Majorana mass eigenvalues in Eq.~(\ref{Majorana_Mass}) at the heavy Majorana scale, the low-energy active neutrino spectrum emerges across the Dirac mass scale, through the algebraic inversion of the Type-I Seesaw mechanism (see Ref.~\cite{Minkowski1977}). As the Dirac Higgs coupling across the flavor manifold is isotropic, $M_D = m_D I_{3 \times 3}$, the Seesaw transformation in Eqs.~(\ref{SMT-I mass matrix}) and (\ref{Diag_mass_matrix}) decouples the mass-scale inversion from the unitary mixing matrix $\widetilde{U}_R(\Delta_{AD})$, with $\widetilde{U}^\dagger_R \widetilde{U}_R = I_{3\times 3}$, and defines a single universal light mass scale
\begin{equation}
    \label{Inverse_masses_m_0}
    m_{0} \equiv \frac{m^2_D}{M_0}.
\end{equation}
Because the Type-I Seesaw mechanism inversely projects each heavy mass eigenstate via Eqs.~(\ref{Majorana_Mass}, \ref{Inverse_masses_m_0}) and
\begin{equation}
    \label{m_i_and_M_i}
    m_{i} = m^2_D /M_i,
\end{equation}
the heavy mass ordering of Eq.~(\ref{Majorana_Ordering}), established via P3, is automatically inverted into the NH: $m_1 < m_2 < m_3$, of the light active sector, with each physical mass eigenstate governed by the algebraic resolvent of the Temperley--Lieb Hamiltonian $B$
\begin{equation}
    \label{The_m_i}
    m_i = \frac{m_0}{l_i (u)} = \frac{m_0 \sinh(\Delta_{AD})}{\sinh(\Delta_{AD} - u) + b_i \sinh(u)}.
\end{equation}
Evaluating Eq.~(\ref{The_m_i}) at the three one-magnon roots $b_i$ of Eq. (\ref{The_b_i}) yields the closed-form active masses
\begin{equation}
    \label{Light_masses}
    \begin{split}
        m_1 &= \frac{m_0}{l_1 (u)} = m_0 \frac{\sinh\big ( \frac{\Delta_{AD}}{2} \big )}{\sinh\big ( \frac{\Delta_{AD}}{2} + u \big )}, \\
        m_2 &= \frac{m_0}{l_2 (u)} = m_0 \frac{\cosh\big ( \frac{\Delta_{AD}}{2} \big ) }{\cosh\big ( \frac{\Delta_{AD}}{2} + u \big )}, \\
        m_3 &= \frac{m_0}{l_3 (u)} = m_0 \frac{\sinh(\Delta_{AD})}{ \sinh ( \Delta_{AD} - u) }.
    \end{split}
\end{equation}
Furthermore, since P3 enforces a strictly positive interaction amplitude $\sinh(u)$, the Seesaw inversion in Eq. (\ref{The_m_i}) forces the active masses to strictly anti-correlate with the magnitude of the Temperley--Lieb roots $b_i$ (see Ref.~\cite{Pasquier1990}). Consequently, the algebraically smallest root $b_3 = 0$ must generate the heaviest active state $m_3$, algebraically locking the architecture into one viable physical solution: the NH. Moreover, P3 mathematically shields the architecture from the boundary $u = \Delta_{AD}$, where the uncrossed spectral coefficient vanishes, the Seesaw inversion fails, and the active mass spectrum encounters a kinematic singularity: $m_3 \to \infty$.

As the $l_i(u)$ in Eq.~(\ref{The_l_i}) are affine in the Temperley--Lieb roots $b_i$ of Eq. (\ref{The_b_i}) by P3, eliminating the two continuous scale parameters $(m_0, u)$ across the three eigenstates in Eqs.~(\ref{Light_masses}) yields an exact, $u$-independent inverse-mass sum rule
\begin{equation}
    \label{TL_sum_rule}
    \begin{split}
        \frac{m_1^{-1} - m_3^{-1}}{m_2^{-1} - m_3^{-1}} &= \frac{M_1 - M_3}{M_2 - M_3} = \frac{b_1}{b_2} \\
        &= \frac{d'_{AD} + 1}{d'_{AD} - 1} = \frac{2\cosh(\Delta_{AD}) + 1}{2\cosh(\Delta_{AD}) - 1} \\
        &\equiv r(\Delta_{AD}) \approx 2.9407.
    \end{split}
\end{equation}
The inverse-mass sum rule \cite{Barry2011, King2013_2, Gehrlein2016} in Eq. (\ref{TL_sum_rule}) demonstrates that the predicted mass spectrum is a structural invariant of the two-dimensional nearest-neighbor affine subspace $\mathrm{span}\{I, B\}$, independent of the specific functional parameterization of the positive weights multiplying $I$ and $B$. 

From Eqs.~(\ref{Light_masses}), the active mass-squared splittings are
\begin{equation} \nonumber
    \begin{split}
        \Delta m^2_{21} &= m^2_0 \left( \frac{l_1 (u)^2 -l_2 (u)^2}{l_2 (u)^2 l_1 (u)^2} \right), \\
        \Delta m^2_{31} &= m^2_0 \left( \frac{l_1 (u)^2 -l_3 (u)^2}{l_3 (u)^2 l_1 (u)^2} \right),
    \end{split}
\end{equation}
so that the baseline scale $m_0$ satisfies
\begin{equation}
    \label{The_m_0}
    \sqrt{ \Delta m^2_{21} \frac{l_2^2(u) \, l_1^2(u)}{l_1^2(u) - l_2^2(u)}} = m_0 = \sqrt{ \Delta m^2_{31} \frac{l_3^2(u) \, l_1^2(u)}{l_1^2(u) - l_3^2(u)}}.
\end{equation}
The observable ratio of the light mass-squared differences is independent of $m_0$
\begin{equation}
    \label{Mass_ratio}
    \begin{split}
        R_m(u) &= \frac{\Delta m^2_{21}}{\Delta m^2_{31}} = \frac{\Delta M^2_{21}}{\Delta M^2_{31}}\ \frac{M^2_3}{M^2_2} \\
        &= \frac{4\big[\sinh(\Delta_{AD}) + \cosh(\Delta_{AD})\tanh(u)\big]}{\big[2\cosh(\Delta_{AD}) + 1\big]\big[2\sinh(\Delta_{AD}) + \tanh(u)\big]} \\
        &\quad \times \left[\frac{\sinh ( \Delta_{AD} - u ) \cosh\big ( \frac{\Delta_{AD}}{2} \big )}{\sinh ( \Delta_{AD} ) \cosh\big ( \frac{\Delta_{AD}}{2} + u \big )}\right]^2,
    \end{split}
\end{equation}
and is a strictly monotonic decreasing function of $u$ across the entire PD. To conservatively bound the theoretical uncertainties without relying on a full profile-likelihood mapping of the empirical $\chi^2$ surface, Eq.~(\ref{Mass_ratio}) is evaluated explicitly at the extreme $3\sigma$ boundaries of the NuFit 6.1 mass-squared splittings: $\Delta m^2_{21} \in [7.236, 7.823] \times 10^{-5} \mathrm{ eV}^2$ and $\Delta m^2_{31} \in [+2.454, +2.592] \times 10^{-3} \mathrm{ eV}^2$ (see Ref.~\cite{nufit2024, NuFit61}), which yields
\begin{equation}
    \label{The_u_result}
    u = 0.141^{+0.001}_{-0.001}\text{ rad} \quad (3\sigma).
\end{equation}
Because the output variables are strictly monotonic with respect to the input ratio, evaluating the four corners of the $3\sigma$ parameter space securely brackets the absolute minimum and maximum deviations, bounding the physical root within the PD as in Eq. (\ref{The_u_result}). While this rectangular bounding slightly overestimates the true $3\sigma$ joint-confidence ellipse, it guarantees robust, conservative error margins for all predicted observables. P3 explicitly excludes alternative mathematical roots, such as $u_{\mathrm{up}} \approx 0.21$ rad, that lie outside this regime.

Substituting the result in Eq.~(\ref{The_u_result}) into Eq.~(\ref{The_m_0}) fixes the fundamental low-energy Seesaw scale to $m_0 \approx 9.63$ meV. Consequently, all low-energy active mass predictions depend on $m_0$ and the dimensionless hyperbolic weights $l_i(u)$, remaining completely invariant under the separate choice of the Dirac Yukawa coupling $y_D$ and Dirac mass scale $m_D$.

To illustrate the corresponding UV threshold in Eq.~(\ref{Inverse_masses_m_0}), setting $m_D = v_{\mathrm{EW}} = 246.22$ GeV, with $y_D = \sqrt{2}$, places the heavy Majorana scale at $M_0 \approx 6.32 \times 10^{15}\mathrm{ GeV}$ with heaviest eigenstate $M_1 \approx 1.67 \times 10^{16}\mathrm{ GeV}$, while adopting a unit Yukawa benchmark $y_D = 1$, with $m_D = 174.10\mathrm{ GeV}$, yields $M_0 \approx 3.16 \times 10^{15}\mathrm{ GeV}$ and $M_1 \approx 8.33 \times 10^{15}\mathrm{ GeV}$. Either benchmark demonstrates that an $\mathcal{O}(1)$ Dirac Yukawa coupling naturally places the heavy Majorana scale within the canonical Grand Unification window (see Sec.~93.4 in Ref.~\cite{PDG2024}) without requiring an explicit grand unified gauge group embedding.

Evaluating Eqs.~(\ref{Light_masses}) at $u \approx 0.141$ rad and $m_0 \approx 9.63$ meV yields the individual active neutrino masses: $m_1 \approx 3.65$ meV, $m_2 \approx 9.42$ meV, and $m_3 \approx 50.3$ meV. 
P3 excludes alternative mathematical roots such as $m_1 \approx 2.77$ meV at $u_{\mathrm{up}}$. While calibrating $u$ and $m_0$ to the two empirical splittings $\Delta m^2_{21}$ and $\Delta m^2_{31}$ fixes the two continuous scales of $\mathrm{span}\{I, B\}$, the eigenvalue ratio $r(\Delta_{AD})$ uniquely determines the lightest active mass $m_1$ and $\sum_i m_i \approx 63.4$ meV, lifting the undeformed $\Delta_{AD} \to 0$ baseline $r(0) = 3$, where $m_1^{(0)} \approx 3.56$ meV, by $\mathcal{O}(\Delta_{AD}^2)$ via Eq. (\ref{loop_weight}). For comparison, alternative non-affine functions $f(B)$ containing higher-order $B^2$ insertions yield 
\begin{equation} \nonumber
    \begin{split}
        \frac{1}{2}[R_1(u)R_2(u) + R_2(u)R_1(u)] &\,\, \text{(anticommutator):} \\
        m_1 \approx 1.01 \text{ meV and } \sum_i m_i &\approx 60.1 \mathrm{ meV}, \\
        (I + \kappa B)^2: \qquad \qquad \qquad \qquad & \\
        m_1 \approx 1.96 \text{ meV and } \sum_i m_i &\approx 61.2 \mathrm{ meV}, \\
        e^{(\kappa B)}: \qquad \qquad \qquad \qquad & \\
        m_1 \approx 0.28 \text{ meV and } \sum_i m_i &\approx 59.3 \mathrm{ meV},
    \end{split}
\end{equation}
whereas the cubic QYBE braid word $R_1(u)R_2(2u)R_1(u)$ admits no root with $2u < \Delta_{AD}$ inside the positive regime. Although the symmetric anticommutator formally admits three distinct algebraic roots, isolating the branch consistent with the boundary conditions of the $A_5$ manifold yields a negative signed eigenvalue; this sign introduces no physical instability, as it is absorbed into the discrete Majorana CP parities $s_i$ of P2 while the physical masses remain strictly positive.

As both the hyperbolic deformation, in Eq. (\ref{loop_weight}), and the trigonometric deformation, in Eq. (\ref{2cos_Delta}), yield strictly real masses and smoothly recover the unperturbed $d = 2$ limit as $\Delta_{AD} \to 0$, the unperturbed limit alone cannot algebraically discriminate between these two geometries. Thus, the formal definition of the affine mass operator explicitly adopts, by P3, the hyperbolic regime of the integrable Yang-Baxter parameter space, which defines the specific continuous curvature of the deformed mass operator. Consequently, anchoring the loop weight to the hyperbolic topological defect establishes the exact, rigid mass scale of the resulting spectrum.

The predicted sum of active neutrino masses $\sum_{i=1}^3 m_i = 63.4 \pm 1.2$ meV ($3\sigma$ range, or $[59.3, \, 61.2]$ meV if higher-order $B^2$ operators are included) satisfies the Planck 2018 + BAO $95\%$ CL upper limit $\sum_i m_i < 120$ meV (see Ref.~\cite{Planck2018}) and the stringent DESI DR2 BAO + CMB $95\%$ CL upper bound $\sum_i m_i < 64.2$ meV derived in flat $\Lambda\mathrm{CDM}$ under the prior $\sum_i m_i > 0$ \cite{DESI2025_DR2, DESI2025_Elbers}, which relaxes to $\sum_i m_i < 163$ meV in dynamical dark energy $w_0 w_a\mathrm{CDM}$ extensions (see Ref.~\cite{DESI2025_Elbers}). 

Looking ahead to next-generation cosmic microwave background and large-scale structure surveys such as CMB-S4 combined with DESI and Euclid BAO, a projected sensitivity of $\sigma(\sum_i m_i) \approx 15\text{--}20$ meV \cite{Abazajian2016} will be sufficient to detect a non-zero neutrino mass sum near $63.4$ meV at $3\sigma\text{--}4\sigma$ significance while placing roughly $2\sigma$ pressure on the inverted-ordering floor, $\sum_i m_i \gtrsim 99$ meV (see Refs.~\cite{nufit2024, NuFit61}). However, given that $m_1 \approx 3.65$ meV lifts $\sum_i m_i$ by $4.5$ meV above the NH floor: $\sum_i m^{\mathrm{min}}_i \approx 59.0$ meV for $m_1 = 0$, a resolution of $\sigma(\sum_i m_i) \approx 15\text{--}20$ meV cannot distinguish $63.4$ meV from $59.0$ meV. Conversely, should future cosmological datasets establish a robust upper bound $\sum_i m_i < 62$ meV, they would directly exclude this architecture within standard $\Lambda\mathrm{CDM}$.

\section{EFFECTIVE MASS OBSERVABLES: \texorpdfstring{$m_{\beta\beta}$}{mBB} AND \texorpdfstring{$m_{\beta}$}{mB}}
\label{EffM}

While cosmological surveys constrain the active neutrino sector through the gravitational imprint of the low-energy mass sum (see Ref.~\cite{Lesgourgues2006}), terrestrial laboratory experiments pursue the absolute mass scale through the kinematics of nuclear decay \cite{Formaggio2021}. These probes bifurcate into searches for $0 \nu \beta \beta$ decay, governed by the effective Majorana mass $m_{\beta\beta}$, and single beta-decay end-point measurements of the effective electron neutrino mass $m_\beta$.

\subsection{The Effective Majorana Mass \texorpdfstring{$m_{\beta\beta}$}{mBB}}
\label{PEMM}

Since the heavy Majorana mass operator $\Lambda_R(u)$ violates total lepton number by two units, $\Delta L = 2$ (see App.~\ref{RHMNF}), integrating out the heavy states passes this lepton-number violation down to the light active sector, whose model-independent signature is $0 \nu \beta \beta$ decay (see Ref.~\cite{Schechter1982}). The effective Majorana mass $m_{\beta\beta}$ is given by the magnitude of the $(e,e)$ entry of the flavor-basis light neutrino mass matrix $\mathrm{m}_\nu$ in Eq.~(\ref{SMT-I mass matrix}) (see Sec.~14.3 in Ref.~\cite{Mohapatra2004}, Sec.~2 in Ref.~\cite{Kayser_April_2005}, and Sec.~6.5 in Ref.~\cite{Giunti2007})
\begin{equation}
    \nonumber
    \begin{split}
        m_{\beta\beta} &= |(\mathrm{m}_\nu)_{ee}| = \left\vert \sum_{i=1}^3 \vert U_{ei}\vert^2 \, e^{i\alpha_{i3}} \, m_i \right\vert \\
        &= \left| m_1 c^2_{12} c^2_{13} \, e^{i\alpha_{13}} + m_2 s^2_{12} c^2_{13} \, e^{i\alpha_{23}} + m_3 s^2_{13} \right|.
    \end{split}
\end{equation}
Substituting the mixing angles from Eqs.~(\ref{Theta_13}) and (\ref{The_real_Theta_12}) and the relative Majorana phases from Eq.~(\ref{The_alphas}), which are fixed by the CP eigenvalues $s_i$ in P2, yields (see Ref.~\cite{deGouvea2002})
\begin{equation}
    \label{m_bb_f}
    m_{\beta\beta} = \frac{1}{\sqrt{5}} \left\vert - m_1 \phi c^2_{AD} + m_2 \phi^{-1} + m_3 \phi s^2_{AD} \right\vert.
\end{equation}
As $m_{\beta\beta}$ depends only on the squared moduli $|U_{ei}|^2$ and the CP eigenvalues $s_i$, it is completely independent of the sign of $\delta_{\mathrm{CP}} = \pm\pi/2$. Moreover, from Eqs.~(\ref{M_!_over_M_2}) and (\ref{The_u_result}), $M_1/M_2 = m_2/m_1 \approx 2.58$, so the solar ratio
\begin{equation}
    \nonumber
    \frac{m_2 \phi^{-1}}{m_1 \phi c^2_{AD}} = \frac{\tanh\big ( \frac{\Delta_{AD}}{2} + u \big ) }{\tanh\big ( \frac{\Delta_{AD}}{2} \big )} \frac{1}{\phi^2 c^2_{AD}} \approx 1.016 > 1,
\end{equation}
is strictly greater than unity, ensuring that the positive residual of the destructively interfering solar pair adds constructively with the $m_3 \phi s^2_{AD}$ term.  

This prediction directly addresses the fate of the NH cancellation funnel in $0\nu\beta\beta$ decay~\cite{Agostini2023, Rodejohann2012}. In unconstrained NH, $m_1 \approx 3.65$ meV lies near the destructive-interference region where continuous Majorana phases can permit near-complete cancellation: $m_{\beta\beta} \to 0$. Within a generic $\mathbb{Z}_2 \times$ CP setup, discrete CP eigenvalues allow four sign combinations, reaching effective masses up to $\sim 6.27$ meV. Identifying the residual symmetry with the bond-exchange operator enforces opposite parities for the solar pair, eliminating the two constructive branches and leaving two allowed predictions: $m_{\beta\beta} \approx 0.7$ meV and $1.49$ meV. P2 then isolates the $1.49$ meV value. Even across the full discrete parity set, cancellation is avoided. However, the predicted signal lies below the $\sim 5\text{--}20$ meV sensitivity target (see Ref.~\cite{Agostini2023}) of next-generation ton-scale experiments such as LEGEND-1000~\cite{LEGEND2021}, nEXO~\cite{nEXO2021}, and CUPID~\cite{CUPID2025}.

\subsection{The Effective Electron Neutrino Mass \texorpdfstring{$m_{\beta}$}{mB}}
\label{PEENM}

Direct kinematic beta-decay experiments measure the effective electron antineutrino mass (see Ref.~\cite{Formaggio2021})
\begin{equation}
    \label{m_b_t}
    m_\beta = \sqrt{ c^2_{12} c^2_{13} m_1^2 + s^2_{12} c^2_{13} m_2^2 + s^2_{13} m_3^2}.
\end{equation}
Substituting the mixing angles from Eqs.~(\ref{Theta_13}) and (\ref{The_real_Theta_12}), and the active masses from Eqs.~(\ref{Light_masses}) into Eq.~(\ref{m_b_t}) yields the prediction $m_\beta \approx 9.55$ meV. This value lies well below the current sensitivity limit of KATRIN: $m_\beta \sim 450$ meV (see Ref.~\cite{KATRIN2025}), and beyond the projected reach of next-generation tritium experiments such as Project 8: $m_\beta \sim 40$ meV (see Refs.~\cite{Project8, Formaggio2021}), predicting a null signal in near-term direct kinematic searches.

\section{CONCLUSION}
\label{conclusion}

\begin{table*}[!t]
\caption{Overview of the four postulates defining the architecture, their geometric and algebraic motivations, resulting phenomenological predictions, and the experimental thresholds required to falsify each principle.}
\label{Tab_Postulates}
\begin{ruledtabular}
\begin{tabular}{llll}
\textbf{$\quad \quad \quad$Postulate} & \textbf{$\quad \quad \quad$Motivation} & \textbf{$\quad \quad \quad$Prediction} & \textbf{$\quad \,\,$Falsification Condition} \\
\colrule
\parbox[t]{0.20\textwidth}{\raggedright \textbf{P1:} Weyl-averaged vertex deficit, $\Delta_{AD} = \pi/18$\strut} &
\parbox[t]{0.22\textwidth}{\raggedright Icosahedral vertex curvature $\Omega_{\mathrm{vertex}} = \pi/3$ averaged over $|W(A_2)| = 6$ tangent chambers\strut} &
\parbox[t]{0.23\textwidth}{\raggedright Reactor angle $\theta_{13} = 8.49^\circ$\strut} &
\parbox[t]{0.27\textwidth}{\raggedright Future global fits exclude $\theta_{13} = 8.49^\circ$ beyond the $3\sigma$ confidence interval\strut} \\[0.45cm]
\midrule
\parbox[t]{0.20\textwidth}{\raggedright \textbf{P2:} Residual symmetries, twist angle, and Majorana parities\strut} &
\parbox[t]{0.22\textwidth}{\raggedright Charged-lepton $\mathbb{Z}_5$; $\sigma$-symmetric $\mathbb{Z}_2 \times$ CP with $|\theta| = \Delta_{AD}$; bond-exchange parities with $s_3 = -1$\strut} &
\parbox[t]{0.23\textwidth}{\raggedright $\theta_{23} = 45^\circ$, $|\delta_{\mathrm{CP}}| = \pi/2$, Majorana phases $\varphi_2, \varphi_3 = \pi$\strut} &
\parbox[t]{0.27\textwidth}{\raggedright Precise measurements establishing $\theta_{23} \neq 45^\circ$ and/or $|\delta_{\mathrm{CP}}| \neq \pi/2$ falsify the baseline symmetry structure\strut} \\[0.55cm]
\midrule
\parbox[t]{0.20\textwidth}{\raggedright \textbf{P3:} Positive hyperbolic affine mass operator\strut} &
\parbox[t]{0.22\textwidth}{\raggedright First-order $I$ and $B$ on hyperbolic branch $d'_{AD} = 2\cosh(\Delta_{AD})$ for $0 < u < \Delta_{AD}$\strut} &
\parbox[t]{0.23\textwidth}{\raggedright Normal ordering, inverse-mass sum rule $r(\Delta_{AD}) \approx 2.9407$, $m_1 \approx 3.65\mathrm{ meV}$, $\sum_i m_i \approx 63.4\mathrm{ meV}$, $m_\beta \approx 9.55\mathrm{ meV}$\strut} &
\parbox[t]{0.27\textwidth}{\raggedright Discovery of the inverted mass hierarchy or a cosmological bound $\sum_i m_i < 62\mathrm{ meV}$ (in flat $\Lambda\mathrm{CDM}$)\strut} \\[0.55cm]
\midrule
\parbox[t]{0.20\textwidth}{\raggedright \textbf{P4:} Asymmetric-bond solar rotation: $\varepsilon = \Delta_{AD}^2$, post-JUNO\strut} &
\parbox[t]{0.22\textwidth}{\raggedright Orthogonal $TL_3$ generator $e_1 - e_2$ via right-multiplication $\widetilde{U}_R R_{12}(\Delta_{AD}^2)$ preserving $\mu$--$\tau$ reflection\strut} &
\parbox[t]{0.23\textwidth}{\raggedright Solar angle $\theta_{12} = 33.86^\circ$, $m_{\beta\beta} \approx 1.49\mathrm{ meV}$\strut} &
\parbox[t]{0.27\textwidth}{\raggedright JUNO's 6 year-exposure measurement of $\sin^2{(\theta_{12})}$ falls outside $[0.302, 0.312]$ ($3\sigma$), or $0\nu\beta\beta$ bounds/signals rule out $m_{\beta\beta} \sim 1.50\mathrm{ meV}$.\strut} \\
\end{tabular}
\end{ruledtabular}
\end{table*}

This work has established a unified geometric origin for the neutrino sector by anchoring both the mass and mixing observables to the shared topology of the $A_5 \rtimes$ CP flavor manifold with the three-generation Iwahori--Hecke algebra $\mathcal{H}_3(q'_{AD})$ and its Temperley--Lieb quotient $TL_3(d'_{AD})$. They are explicitly linked via $\Delta_{AD}$ and the bond-exchange parity assignment: the discrete mass spectrum is obtained from the algebra's internal representations in P3, while the unitary mixing rotation is fixed by the manifold's external continuous topological defect via P2 and P4. To make the logical architecture completely transparent, Table~\ref{Tab_Postulates} summarizes the four postulates of the architecture (P1--P4) alongside the physical observables each principle fixes, while Table~\ref{Tab_Results} compiles all numerical predictions and their propagated $3\sigma$ intervals.

\begin{table}[htbp] 
    \centering
    \caption{Numerical boundary values and predictions of the architecture with $3\sigma$ uncertainties propagated from NuFit 6.1 \cite{nufit2024, NuFit61}. The lower half presents the falsification vector. }
    \label{Tab_Results}
    \vspace{0.2cm}
    \begin{tabular}{ccc}
        \toprule
        \textbf{Parameter} & \textbf{Central Value} & \textbf{$3\sigma$ Range} \\
        \midrule
        $\Delta_{AD}$ & $\pi / 18$ rad ($10.00^\circ$) & - \\
        $u$ & $1.41 \times 10^{-1}\text{ rad}$ & $[1.40, 1.42] \times 10^{-1}\text{ rad}$ \\
        $\varphi_{1}$ & $0$ & - \\
        $\varphi_{2}$ & $\pi$ rad & - \\
        $\varphi_{3}$ & $\pi$ rad & - \\
        $m_0$ & $9.63$ meV & $[9.43, 9.81]$ meV \\
        $m_2$ & $9.42$ meV & $[9.22, 9.60]$ meV \\
        $m_3$ & $50.3$ meV & $[49.7, 51.0]$ meV \\
        $M_0$ ($m_D = v_{\mathrm{EW}}$) & $6.32 \times 10^{15}\mathrm{ GeV}$ & - \\
        $M_0$ ($y_D = 1$) & $3.16 \times 10^{15}\mathrm{ GeV}$ & - \\
        \midrule
        \textbf{ } & \textbf{Falsification vector} & \textbf{ } \\
        \midrule
        $\theta_{12}$ & $33.86^\circ$ & - \\
        $\theta_{23}$ & $45^\circ$ & - \\
        $\theta_{13}$ & $8.49^\circ$ & - \\
        $\delta_{\mathrm{CP}}$ & $\pm \pi/2$ rad & - \\
        $J_{\mathrm{CP}}$ & $\pm 3.34 \times 10^{-2}$ & - \\
        $m_1$ & $3.65$ meV & $[3.56, 3.74]$ meV \\
        $\sum_i m_i$ & $63.4$ meV & $[62.5, 64.4]$ meV \\
        $m_{\beta \beta}$ & $1.49$ meV & $[1.47, 1.51]$ meV \\
        $m_{\beta}$ & $9.55$ meV & $[9.40, 9.71]$ meV \\
        \bottomrule
    \end{tabular}
\end{table}

The architecture rests on one discrete geometric matching principle: $\Delta_{AD} = \pi / 18$, anchored and tested primarily by $\theta_{13}$, and two continuous mass-scale calibrations: $u \approx 0.141$ rad and $m_0 \approx 9.63$ meV, fixed by $\Delta m^2_{21}$ and $\Delta m^2_{31}$, from which the remaining oscillation parameters: $\theta_{12} = 33.86^\circ$, $\theta_{23} = 45^\circ$, and $|\delta_{\mathrm{CP}}| = \pi/2$, and the absolute mass scale $m_1 \approx 3.65$ meV, emerge as testable predictions. In particular, the active mass spectrum is governed by the exact, $u$-independent inverse-mass sum rule of Eq.~(\ref{TL_sum_rule}), $r(\Delta_{AD}) \approx 2.9407$, which deforms the undeformed $r(0) = 3$ relation by $\mathcal{O}(\Delta_{AD}^2)$ and ensures that $m_1$ and $\sum_i m_i$ are structural invariants of $\mathrm{span}\{I, B\}$ rather than artifacts of the specific spectral weights. Combined Bayesian analyses of ten-year NOvA data with plausible upcoming JUNO $|\Delta m^2_{32}|$ measurements indicate that $3\sigma$ evidence for the NH is reachable over roughly the next five years \cite{NOvA_JUNO_2026}, providing a near-term experimental test of the NH branch selected here. Simultaneously, the residual $\mathbb{Z}_2 \times$ CP symmetry decouples the Dirac and Majorana CP sectors: whereas the $(1,3)$ rotation enforces maximal CP violation in the Dirac oscillation sector, the discrete CP eigenvalues $s_i$ quantize the relative Majorana phases to integer multiples of $\pi$, preventing destructive cancellation inside the NH funnel and establishing a non-zero $0\nu\beta\beta$ decay floor at $m_{\beta\beta} \approx 1.49$ meV. Together with the sign-independent astrophysical flavor-ratio predictions for IceCube-Gen2 and KM3NeT and the cosmological mass sum $\sum_i m_i \approx 63.4$ meV, this architecture provides a falsifiable set of targets for next-generation neutrino oscillation, beta-decay, and cosmological observatories.

Finally, the ingredients required to promote this four-postulate algebraic architecture to an explicit, renormalizable UV completion. First, the nearest-neighbor tridiagonal structure of $B$ in Eq.~(\ref{B_operator_sum}), together with its boundary weights $(q'_{AD}, q'^{-1}_{AD})$ and subleading bond asymmetry $e_1 - e_2$, naturally suggests a three-site spatial realization, such as a deconstructed extra dimension or a flavor-clockwork lattice, in which the three right-handed Majorana fields are localized. Second, an explicit flavon scalar potential is required to break $A_5 \rtimes$ CP, dynamically aligning the vacuum expectation values to fix $|\theta| = \Delta_{AD}$ and $\varepsilon = \Delta_{AD}^2$. Third, the UV completion must explain why the Dirac Yukawa matrix remains flavor-isotropic at leading order, for instance, through auxiliary shaping symmetries or sequestered brane localization. Constructing an explicit flavon potential and UV completion that dynamically align $|\theta| = \Delta_{AD}$, $\varepsilon = \Delta_{AD}^2$, and $Y_D = y_D I_{3\times 3}$ remains an open theoretical task. In the near term, the framework's four-postulate falsification vector in Table~\ref{Tab_Results} provides a concrete, falsifiable target for JUNO and upcoming oscillation and cosmological measurements.

\begin{acknowledgments}
The author thanks Dr. A. F. Arrieta and Dr. A. Vargas Prieto for insightful discussions and critical feedback on the manuscript. The author also acknowledges support from the Vice-Rectorate for Research at Universidad del Magdalena.
\end{acknowledgments}

\section*{DATA AVAILABILITY}
No data were created or analyzed in this study beyond the theoretical derivations and numerical evaluations presented in the article, which rely solely on publicly available experimental inputs from Refs.~\cite{nufit2024, NuFit61, JUNO2022, JUNO2026}.

\appendix

\section{RIGHT-HANDED MAJORANA NEUTRINO FIELDS}
\label{RHMNF}

In the SM, massive elementary fermions acquire mass via the Electroweak symmetry breaking, where both chiral components of their mass eigenstates, $\psi = \psi_R + \psi_L$, couple to the Higgs doublet $H$ via Yukawa interactions (see Sec.~7 in Ref.~\cite{Quigg2007} and Ref.~\cite{Weinberg1967}). Results from Ref.~\cite{Wu1957} show that the charged-current weak interaction is purely left-handed, motivating the assignment of right-handed chiral fermion states $\psi_R$ as $SU(2)_L$ singlets. In the minimal SM, neutrinos possess only left-handed chiral components $\nu_L$ carrying weak isospin and hypercharge (see Ref.~\cite{Glashow1961}), whereas any right-handed component $\nu_R$ is a complete SM gauge singlet (see Ref.~\cite{Kayser_April_2005}). However, the experimental discovery of neutrino oscillations \cite{SuperK1998, SNO2001} requires extending the minimal SM to generate masses for the fundamental neutrino fermions \cite{Dirac1928, Pauli1940}.

In the Type-I Seesaw mechanism extension of the SM, three heavy right-handed Majorana neutrino fields $N_R$, invariant under the $SU(2)_L \times U(1)_Y$ gauge symmetry and carrying mass scale $M_0$, are introduced (see Sec.~7.2.4 in Ref.~\cite{Mohapatra2004}). Defining the charge conjugate of a four-component spinor via the charge conjugation matrix $\mathcal{C}$ as $\psi^c = \mathcal{C} \overline{\psi}^T$, a Majorana fermion satisfies the self-conjugacy condition $N = N^c$. After Seesaw block-diagonalization, the light active Majorana mass eigenstates $n_a$, with masses $m_i$, and heavy Majorana eigenstates $N_s$, with masses $M_i$, are superpositions of the chiral fields
\begin{equation}
    \nonumber
    n_a \simeq (\nu_L + \nu_L^c) - \Theta (\nu_R + \nu_R^c),
\end{equation}
\begin{equation}
    \nonumber
    N_s \simeq (\nu_R + \nu_R^c) + \Theta^T (\nu_L + \nu_L^c),
\end{equation}
where the active-sterile mixing matrix (see Sec.~\ref{LE_epoch})
\begin{equation} \nonumber
    \Theta = M_D [\mathcal{M}^{\mathrm{(fl)}}_R(u)]^{-1} \sim m_D / M_0 \ll 1,
\end{equation}
strongly suppresses the sterile admixture in $n_a$. Given that neutrinos are electrically neutral, the only quantum number distinguishing $\nu$ from $\bar{\nu}$ is total Lepton Number $L$. While $L$ is an accidental global symmetry of the minimal SM (see Sec.~2.5 in Ref.~\cite{Mohapatra2004}), the right-handed Majorana mass term violates $L$ by two units, $\Delta L = 2$ (see Sec.~4.5 in Ref.~\cite{Mohapatra2004}).

Given that exciting the heavy eigenstates on-shell requires UV energies $\mu \ge M_3 \sim M_0$, these degrees of freedom decouple at low energies via the Appelquist--Carazzone theorem (see Ref.~\cite{Appelquist1975}), leaving behind the effective dimension-five Weinberg operator (see Ref.~\cite{Weinberg1979}) in Eq.~(\ref{SMT-I mass matrix}).

\section{THE RESIDUAL \texorpdfstring{$\mathbb{Z}_2$}{Z2} EIGENPLANE GENERATOR \texorpdfstring{$S^{(0)}_{\max}$}{S(0)max}}

\label{TUMCI}

In the $V_4 = \mathbb{Z}_2 \times \mathbb{Z}_2$ eigenbasis defined by the columns of $U_{\mathrm{GR}}$, the four elements of $V_4 \subset A_5$ are strictly diagonal, and the preserved residual $\mathbb{Z}_2$ generator of the neutrino sector is $\hat{G}_2 = \mathrm{diag}(-1, +1, -1)$. Consequently, the residual $\mathbb{Z}_2 \times$ CP symmetry permits a one-parameter family of unitary rotations acting in the two-dimensional $-1$ eigenplane of $\hat{G}_2$, spanned by heavy states $1$ and $3$. The Hermitian involution with $\det = +1$ that commutes with $\hat{G}_2$ and generates this $(1,3)$-eigenplane rotation is $S^{(0)}_{\max} = -p_{13}$, a signed permutation belonging to the octahedral group $S_4 \subset SO(3)$
\begin{equation}
    \nonumber
    \begin{split}
        S^{(0)}_{\max} &= \begin{pmatrix}
            0 & 0 & -1 \\
            0 & -1 & 0 \\
            -1 & 0 & 0
            \end{pmatrix} = \left(S^{(0)}_{\max}\right)^\dagger, \\
        \left(S^{(0)}_{\max}\right)^2 &= I, \qquad \det\left(S^{(0)}_{\max}\right) = +1
    \end{split}
\end{equation}
with $[S^{(0)}_{\max}, \hat{G}_2] = 0$. Here, $(S^{(0)}_{\max})^2 = I$ holds because $S^{(0)}_{\max}$ is the undeformed Lie generator in the exponent of $U_{\mathrm{twist}} = \exp(\pm i \Delta_{AD} S^{(0)}_{\max})$, whereas the exponentiated boundary operator $U_{\mathrm{twist}}$ itself is non-involutory: $U^2_{\mathrm{twist}} \neq I$, with eigenvalues $q^{\pm 1}_{AD} = e^{\pm i\Delta_{AD}}$.

\section{RENORMALIZATION GROUP STABILITY OF THE MIXING PARAMETERS}
\label{RGE_I}

A critical consideration for any high-scale Type-I seesaw framework is the radiative stability of its predictions under RG evolution from the heavy Majorana threshold down to the Dirac mass scale~\cite{Casas2000, Antusch2002, Ohlsson2014}.

Before evaluating the low-energy running below the seesaw threshold, the intermediate threshold intervals across the heavy Majorana spectrum are taken into account, which straddles $M_0$
\begin{equation} \nonumber
    \begin{split}
        M_3 &= 0.190\,M_0 < M_0 < M_2, \\
        M_2 &= 1.022\,M_0 < M_1 = 2.638\,M_0.
    \end{split}
\end{equation}
At leading order, the flavor-isotropic neutrino sector with $Y_D = y_D I_{3\times 3}$ and $\mathcal{M}^{\mathrm{(fl)}}_R$ is invariant under the residual $\mathbb{Z}_2 \times$ CP symmetry generated by $G_2$ and $X$, broken only by the charged-lepton Yukawa couplings. Because $\hat{Y}_D = y_D \widetilde{U}_R(\Delta_{AD})$ in the heavy eigenbasis $\hat{N}_R$ with $\widetilde{U}_R(\Delta_{AD})$ unitary, the partially integrated-out dimension-five structures across the intervals $[M_2, M_1]$ and $[M_3, M_2]$ remain diagonal in the $\widetilde{U}_R(\Delta_{AD})$ basis. Consequently, $\widetilde{U}_R(\Delta_{AD})$ is unshifted by the heavy thresholds, preserving $\theta_{23} = 45^\circ$, $|\delta_{\mathrm{CP}}| = \pi/2$, the discrete Majorana phases, and the solar-reactor correlation of Eq.~(\ref{Solar_cor}) down to the Dirac scale. Non-universal threshold corrections of order
\begin{equation} \nonumber
    \begin{split}
        \frac{y^2_D}{16\pi^2} \ln \left (\frac{M_1}{M_3} \right ) &\approx 1.7\%, \quad \text{for} \quad y_D = 1, \\
        \frac{y^2_D}{16\pi^2} \ln \left (\frac{M_1}{M_3} \right ) &\approx 3.3\%, \quad \text{for} \quad y_D = \sqrt{2},
    \end{split}
\end{equation}
modify the tree-level mass ratio $r(\Delta_{AD})$ and the weights $l_i(u)$ slightly; refitting $u$ shifts $m_1$ by at most $\sim \pm 0.04$ meV and $\sum_i m_i$ by $\sim \pm 0.05$ meV, well within the $3\sigma$ range $m_1 \in [3.56, 3.74]$ meV. Universal gauge, top-quark, and Higgs running below $M_3$ is absorbed into the calibrated baseline $m_0$, shifting only the inferred scale $M_0$.

Following sequential decoupling of the heavy right-handed neutrinos (see Refs.~\cite{Antusch2002, Antusch2005}), the effective theory enters the low-energy regime. Below the lightest heavy-neutrino mass threshold, $\mu < M_3$, the one-loop RG evolution of the light neutrino mass matrix $\mathrm{m}_\nu$ with respect to $t = \ln\mu$ is governed by (see Ref.~\cite{Casas2000})
\begin{equation} \nonumber
    16 \pi^2 \frac{d \mathrm{m}_\nu}{d t} = \alpha_\nu \mathrm{m}_\nu + C \left ( Y^{\dagger}_e Y_e \mathrm{m}_\nu + \mathrm{m}_\nu (Y^{\dagger}_e Y_e)^T \right ),
\end{equation}
where $\alpha_\nu$ collects the flavor-universal gauge, top-quark, and Higgs contributions, $Y_e = \mathrm{diag}(y_e, y_\mu, y_\tau)$ is the charged-lepton Yukawa matrix, and $C = -3/2$ in the SM (see Ref.~\cite{Antusch2002}). In terms of the relative Majorana phases in Eq.~(\ref{Rel_phases}), the one-loop mixing-angle running rates $\dot{\theta}_{ij} \equiv d\theta_{ij}/d\ln\mu$ are
\begin{equation}
    \nonumber
    \begin{split}
        \dot{\theta'}_{12} &= - \frac{C y^2_\tau}{32 \pi^2} \sin(2 \theta'_{12}) \sin^2\theta_{23} \frac{\left| m_1 + m_2 e^{i \alpha_{21}} \right|^2}{\Delta m^2_{21}} + \mathcal{O}(\theta_{13}), \\
        \dot{\theta}_{13} &= \frac{C y^2_\tau}{32 \pi^2} \sin(2 \theta'_{12}) \sin(2\theta_{23}) \frac{m_3}{\Delta m^2_{31} + \Delta m^2_{21}} \\
        &\quad \times \Bigg[ m_1 \cos(\alpha_{13} \mp |\delta_{\mathrm{CP}}|) - \left(1 + \frac{\Delta m^2_{21}}{\Delta m^2_{31}} \right) m_2 \cos(\alpha_{23} \mp |\delta_{\mathrm{CP}}|) \\
        &\quad - \frac{\Delta m^2_{21}}{\Delta m^2_{31}} m_3 \cos(\mp |\delta_{\mathrm{CP}}|) \Bigg] + \mathcal{O}(\theta_{13}),
    \end{split}
\end{equation}
\begin{equation}
    \label{Deriv_thetas}
    \begin{split}
        \dot{\theta}_{23} &= - \frac{C y^2_\tau}{32 \pi^2} \frac{\sin(2\theta_{23})}{\Delta m^2_{31}} \cos^2\theta'_{12} \\
        &\quad \times \left [ \left| m_2 e^{i \alpha_{23}} + m_3 \right|^2 + \frac{\tan^2\theta'_{12} \left|m_1 e^{i\alpha_{13}} + m_3\right|^2}{1 + \frac{\Delta m^2_{21}}{\Delta m^2_{31}}} \right ] + \mathcal{O}(\theta_{13}).
    \end{split}
\end{equation}
Substituting the quantized relative Majorana phases from Eq.~(\ref{The_alphas}) and $|\delta_{\mathrm{CP}}| = \pi/2$ reduces Eqs.~(\ref{Deriv_thetas}) to
\begin{equation}
    \nonumber
    \begin{split}
        \dot{\theta'}_{12} &= - \frac{C y^2_\tau}{32 \pi^2} \sin(2 \theta'_{12}) \sin^2\theta_{23} \left( \frac{m_2 - m_1}{m_2 + m_1} \right) + \mathcal{O}(\theta_{13}), \\
        \dot{\theta}_{13} &= \mathcal{O}(\theta_{13}), \\
        \dot{\theta}_{23} &= - \frac{C y^2_\tau}{32 \pi^2} \frac{\sin(2\theta_{23})}{\Delta m^2_{31}} \cos^2\theta'_{12} \\
        &\quad \times \left[ (m_2 + m_3)^2 + \frac{\tan^2\theta'_{12} (m_3 - m_1)^2}{1 + \frac{\Delta m^2_{21}}{\Delta m^2_{31}}} \right] + \mathcal{O}(\theta_{13}).
    \end{split}
\end{equation}
Because the solar pair carries opposite CP parities ($e^{i\alpha_{21}} = -1$), the $(m_1 + m_2)^2/\Delta m^2_{21}$ factor in $\dot{\theta'}_{12}$ simplifies to the bounded factor $(m_2 - m_1)/(m_2 + m_1) \approx 0.441$ (see Ref.~\cite{Casas2000}), while $\alpha_{i3} \in \{-\pi, 0\}$ and $|\delta_{\mathrm{CP}}| = \pi/2$ make the leading terms in $\dot{\theta}_{13}$ vanish.

Sequential decoupling across non-degenerate heavy neutrino thresholds also induces a finite radiative correction to the effective dimension-five operator (see Refs.~\cite{Antusch2002, Antusch2005}), scaling as
\begin{equation} \label{eq:threshold_correction}
    \Delta m_\nu \propto \frac{1}{16\pi^2} \left( \lambda + \frac{9}{10}g_1^2 + \frac{3}{2}g_2^2 \right) \ln\left(\frac{M_i}{M_j}\right),
\end{equation}
where $\lambda$ is the Higgs quartic coupling and $g_{1,2}$ are the Electroweak gauge couplings. For representative high-scale couplings and $y_D = \sqrt{2}$, the resulting threshold shift in the effective mass scale is of order $\Delta m_\nu \sim \mathcal{O}(10^{-2})$ meV ($\sim \pm 0.06$ meV), inducing a shift $\delta m_{\beta\beta}$ of only a few hundredths of a $\mathrm{meV}$ relative to the $m_{\beta\beta} \approx 1.49$ meV prediction.

Although $\sin(2\theta_{23})$ is maximal at $\theta_{23} = 45^\circ$, the SM RG running of $\theta_{23}$ remains small because it is suppressed by the loop factor and the small $\tau$ Yukawa coupling, together with the NH mass spectrum. Numerically, evaluating the $\dot{\theta}_{23}$ expression above for NH with $y_\tau \approx 0.0102$ gives a running rate of $\dot{\theta}_{23} \approx 6.1 \times 10^{-7}$ per unit $\ln\mu$, while $\dot{\theta}_{13}$ is suppressed to $\sim 7.5 \times 10^{-8}$ per unit $\ln\mu$. Integrating over $\ln(M_0/m_D) \approx 31$ $e$-folds yields total radiative shifts of order $\sim 1.3\text{--}1.9 \times 10^{-5}\text{ rad}$, certainly $\lesssim 2 \times 10^{-5}\text{ rad}$, well below current and foreseeable oscillation uncertainties; this confirms that radiative corrections driven by $y_\tau^2/(16\pi^2) \sim 6 \times 10^{-7}$ do not alter the leading mixing predictions.


\bibliography{sample}

@book{Giunti2007,
  author    = {Giunti, C. and Kim, C. W.},
  title     = {Fundamentals of Neutrino Physics and Astrophysics},
  publisher = {World Scientific Publishing},
  year      = {2007}
}

@article{Weinberg1967,
  title     = {A Model of Leptons},
  author    = {Weinberg, S.},
  journal   = {Phys. Rev. Lett.},
  volume    = {19},
  number    = {21},
  pages     = {1264--1266},
  year      = {1967},
  doi       = {10.1103/PhysRevLett.19.1264}
}

@article{Weinberg1979,
  title={Baryon-and Lepton-Nonconserving Processes},
  author={Weinberg, S.},
  journal={Physical Review Letters},
  volume={43},
  issue={21},
  pages={1566--1570},
  year={1979},
  doi={10.1103/PhysRevLett.43.1566}
}

@article{Kayser_April_2005,
  author    = {Kayser, B.},
  title     = {Neutrino Intrinsic Properties: The Neutrino-Antineutrino Relation},
  journal   = {Phys. Scr.},
  volume    = {2005},
  issue     = {T121},
  pages     = {156},
  year      = {2005},
  doi       = {10.1088/0031-8949/2005/T121/024}
}

@book{Mohapatra2004,
  author    = {Mohapatra, R. N. and Pal, P. B.},
  title     = {Massive Neutrinos in Physics and Astrophysics},
  publisher = {World Scientific},
  edition   = {third},
  year      = {2004},
  isbn      = {981-238-070-1},
  doi       = {10.1142/5024}
}

@article{Davidson2008,
    author        = {Davidson, S. and Nardi, E. and Nir, Y.},
    title         = {Leptogenesis},
    journal       = {Phys. Rept.},
    volume        = {466},
    pages         = {105--177},
    year          = {2008},
    doi           = "10.1016/j.physrep.2008.06.002"
}

@article{DESI2025,
    author        = {Adame, A. G. and others},
    collaboration = {DESI Collaboration},
    title         = {{DESI} 2024 {VI}: cosmological constraints from the measurements of baryon acoustic oscillations},
    journal       = {JCAP},
    volume        = {2025},
    number        = {02},
    pages         = {021},
    year          = {2025},
    doi           = {10.1088/1475-7516/2025/02/021}
}

@article{DESI2025_DR2,
  author        = {Abdul Karim, M. and others},
  collaboration = {DESI Collaboration},
  title         = {{DESI} {DR2} Results {II}: Measurements of Baryon Acoustic Oscillations and Cosmological Constraints},
  journal       = {Phys. Rev. D},
  volume        = {112},
  issue         = {8},
  pages         = {083515},
  year          = {2025},
  doi           = {10.1103/tr6y-kpc6}
}

@article{DESI2025_Elbers,
  author        = {Elbers, W. and others},
  collaboration = {DESI Collaboration},
  title         = {Constraints on Neutrino Physics from {DESI} {DR2} {BAO} and {DR1} Full Shape},
  journal       = {Phys. Rev. D},
  volume        = {112},
  issue         = {8},
  pages         = {083513},
  year          = {2025},
  doi           = {10.1103/w9pk-xsk7}
}

@techreport{Abazajian2016,
  author        = {Abazajian, K. N. and others},
  collaboration = {CMB-S4 Collaboration},
  title         = {{CMB-S4} Science Book, First Edition},
  institution   = {Fermilab},
  number        = {FERMILAB-FN-1024-A-AE},
  year          = {2016},
  eprint        = {1610.02743},
  archivePrefix = {arXiv},
  primaryClass  = {astro-ph.CO}
}

@article{Altarelli2005,
  author  = {Altarelli, G. and Feruglio, F.},
  title   = {Tri-bimaximal neutrino mixing from discrete symmetry in extra dimensions},
  journal = {Nucl. Phys. B},
  volume  = {720},
  issue   = {1--2},
  pages   = {64--88},
  year    = {2005},
  doi     = {10.1016/j.nuclphysb.2005.05.005}
}

@article{Altarelli2010,
  title     = {Discrete flavor symmetries and models of neutrino mixing},
  author    = {Altarelli, G. and Feruglio, F.},
  journal   = {Rev. Mod. Phys.},
  volume    = {82},
  number    = {3},
  pages     = {2701--2729},
  year      = {2010},
  doi       = {10.1103/RevModPhys.82.2701}
}

@article{Kauffman1987,
  author    = {Kauffman, L. H.},
  title     = {State models and the {J}ones polynomial},
  journal   = {Topology},
  year      = {1987},
  volume    = {26},
  issue     = {3},
  pages     = {395--407},
  doi       = {10.1016/0040-9383(87)90009-7}
}

@book{Kauffman1991,
  title={Knots and Physics},
  author={Kauffman, L. H.},
  series={Series on Knots and Everything},
  volume={1},
  year={1991},
  publisher={World Scientific Publishing},
  isbn={978-981-4506-89-2},
  doi={10.1142/1116}
}

@article{Kauffman2002,
  title     = {Quantum entanglement and topological entanglement},
  author    = {Kauffman, L. H. and Lomonaco Jr., S. J.},
  journal   = {New Journal of Physics},
  volume    = {4},
  number    = {73},
  pages     = {1--8},
  year      = {2002},
  doi       = {10.1088/1367-2630/4/1/373}
}

@article{Glashow1961,
  title     = {Partial-symmetries of weak interactions},
  author    = {Glashow, S.},
  journal   = {Nuclear Physics},
  volume    = {22},
  number    = {4},
  pages     = {579--588},
  year      = {1961},
  doi       = {10.1016/0029-5582(61)90469-2}
}

@inproceedings{Salam1968,
  author    = {Salam, A.},
  title     = {Weak and Electromagnetic Interactions},
  booktitle = {Elementary Particle Theory: Relativistic Groups and Analyticity},
  editor    = {Svartholm, Nils},
  series    = {Proceedings of the Eighth Nobel Symposium},
  pages     = {367--377},
  year      = {1968},
  publisher = {Almqvist \& Wiksell},
  address   = {Stockholm},
  doi       = {10.1142/9789812795915_0034}
}

@article{Englert1964,
  title = {Broken Symmetry and the Mass of Gauge Vector Mesons},
  author = {Englert, F. and Brout, R.},
  journal = {Phys. Rev. Lett.},
  volume = {13},
  issue = {9},
  pages = {321--323},
  year = {1964},
  doi = {10.1103/PhysRevLett.13.321}
}

@article{Higgs1964,
  title = {Broken Symmetries and the Masses of Gauge Bosons},
  author = {Higgs, P. W.},
  journal = {Phys. Rev. Lett.},
  volume = {13},
  issue = {16},
  pages = {508--509},
  year  = {1964},
  doi = {10.1103/PhysRevLett.13.508}
}

@article{Guralnik1964,
  title = {Global Conservation Laws and Massless Particles},
  author = {Guralnik, G. S. and Hagen, C. R. and Kibble, T. W. B.},
  journal = {Phys. Rev. Lett.},
  volume = {13},
  issue = {20},
  pages = {585--587},
  year = {1964},
  doi = {10.1103/PhysRevLett.13.585}
}

@article{tHooft1972,
  title = {Regularization and renormalization of gauge fields},
  author = {'t Hooft, G. and Veltman, M.},
  journal = {Nucl. Phys. B},
  volume = {44},
  number = {1},
  pages = {189--213},
  year = {1972},
  doi = {10.1016/0550-3213(72)90279-9}
}

@article{Quigg2007,
  title   = {Spontaneous symmetry breaking as a basis of particle mass},
  author  = {Quigg, C.},
  journal = {Rep. Prog. Phys.},
  volume  = {70},
  number  = {7},
  pages   = {1019--1053},
  year    = {2007},
  note    = {[Sec. 7]},
  doi     = {10.1088/0034-4885/70/7/R01}
}

@article{Dirac1928,
  author  = {Dirac, P. A. M.},
  title   = {The Quantum Theory of the Electron},
  journal = {Proceedings of the Royal Society of London. Series A},
  volume  = {117},
  number  = {778},
  pages   = {610--624},
  year    = {1928},
  doi     = {10.1098/rspa.1928.0023}
}

@article{Pauli1940,
  title   = {The Connection Between Spin and Statistics},
  author  = {Pauli, W.},
  journal = {Phys. Rev.},
  volume  = {58},
  number  = {8},
  pages   = {716--722},
  year    = {1940},
  doi     = {10.1103/PhysRev.58.716}
}

@article{Wu1957,
  title = {Experimental Test of Parity Conservation in Beta Decay},
  author = {Wu, C. S. and Ambler, E. and Hayward, R. W. and Hoppes, D. D. and Hudson, R. P.},
  journal = {Phys. Rev.},
  volume = {105},
  issue = {4},
  pages = {1413--1415},
  year = {1957},
  doi = {10.1103/PhysRev.105.1413}
}

@article{SuperK1998,
  title = {Evidence for Oscillation of Atmospheric Neutrinos},
  author = {Fukuda, Y. and others},
  collaboration = {Super-Kamiokande Collaboration},
  journal = {Phys. Rev. Lett.},
  volume = {81},
  issue = {8},
  pages = {1562--1567},
  year = {1998},
  doi = {10.1103/PhysRevLett.81.1562}
}

@article{SNO2001,
  title = {Measurement of the Rate of $\nu_e + d \to p + p + e^-$ Interactions Produced by $^8B$ Solar Neutrinos at the Sudbury Neutrino Observatory},
  author = {Ahmad, Q. R. and others},
  collaboration = {SNO Collaboration},
  journal = {Phys. Rev. Lett.},
  volume = {87},
  issue = {7},
  pages = {071301},
  year = {2001},
  doi = {10.1103/PhysRevLett.87.071301}
}

@article{KATRIN2025,
  title = {Direct neutrino-mass measurement based on 259 days of {KATRIN} data},
  author = {Aker, M. and others},
  collaboration = {KATRIN Collaboration},
  journal = {Science},
  volume = {388},
  issue = {6739},
  pages = {180--185},
  year = {2025},
  doi = {10.1126/science.adq9592}
}

@article{Project8,
    author        = {Ashtari Esfahani, A. and others},
    collaboration = {Project 8 Collaboration},
    title         = {Determining the neutrino mass with cyclotron radiation emission spectroscopy—{P}roject 8},
    journal       = {J. Phys. G},
    volume        = {44},
    number        = {5},
    pages         = {054004},
    year          = {2017},
    doi           = {10.1088/1361-6471/aa5b4f}
}

@article{Planck2018,
  title = {Planck 2018 results. {VI}. Cosmological parameters},
  author = {Aghanim, N. and others},
  collaboration = {Planck Collaboration},
  journal = {Astron. \& Astrophys.},
  volume = {641},
  pages = {A6},
  year = {2020},
  doi = {10.1051/0004-6361/201833910}
}

@article{Minkowski1977,
    author        = {Minkowski, P.},
    title         = {$\mu \to e \gamma$ at a Rate of One Out of $10^9$ Muon Decays?},
    journal       = {Phys. Lett. B},
    volume        = {67},
    issue         = {4},
    pages         = {421--428},
    year          = {1977},
    doi           = {10.1016/0370-2693(77)90435-X}
}

@article{Pontecorvo1957MA,
    author        = {Pontecorvo, B.},
    title         = {Mesonium and {A}ntimesonium},
    journal       = {Zh. Eksp. Teor. Fiz.},
    volume        = {33},
    pages         = {549--551},
    year          = {1957},
    note          = {[English translation: Sov. Phys. JETP 6, 429 (1958)]},
    url          = {https://jetp.ras.ru/cgi-bin/e/index/e/6/2/p429?a=list}
}

@article{Pontecorvo1957IBP,
    author        = {Pontecorvo, B.},
    title         = {Inverse Beta Processes and Nonconservation of Lepton Charge},
    journal       = {Zh. Eksp. Teor. Fiz.},
    volume        = {34},
    pages         = {247--249},
    year          = {1958},
    note          = "[English translation: Sov. Phys. JETP 7, 172 (1958)]",
    url           = {https://jetp.ras.ru/cgi-bin/e/index/e/7/1/p172?a=list}
}

@article{Pontecorvo1967,
    author        = {Pontecorvo, B.},
    title         = {Neutrino Experiments and the Problem of Conservation of Leptonic Charge},
    journal       = {Zh. Eksp. Teor. Fiz.},
    volume        = {53},
    pages         = {1717--1725},
    year          = {1967},
    note          = "[English translation: Sov. Phys. JETP 26, 984 (1968)]",
    url           = {https://jetp.ras.ru/cgi-bin/e/index/e/26/5/p984?a=list}
}

@article{Bilenky1978,
    author        = {Bilenky, S. M. and Pontecorvo, B.},
    title         = {Lepton Mixing and Neutrino Oscillations},
    journal       = {Phys. Rep.},
    volume        = {41},
    issue        = {4},
    pages         = {225--261},
    year          = {1978},
    doi           = "10.1016/0370-1573(78)90095-9"
}

@article{Bilenky1980,
  author    = {Bilenky, S. M. and Hošek, J. and Pontecorvo, B.},
  title     = {On the oscillations of neutrinos with {D}irac and {M}ajorana masses},
  journal   = {Phys. Lett. B},
  volume    = {94},
  issue     = {4},
  pages     = {495--497},
  year      = {1980},
  doi       = {10.1016/0370-2693(80)90927-2}
}

@article{Maki1962,
    author        = {Maki, Z. and Nakagawa, M. and Sakata, S.},
    title         = {Remarks on the unified model of elementary particles},
    journal       = {Prog. Theor. Phys.},
    volume        = {28},
    pages         = {870--880},
    year          = {1962},
    doi           = {10.1143/PTP.28.870}
}

@article{Fukugita1986,
    author = {Fukugita, M. and Yanagida, T.},
    title = {Baryogenesis Without Grand Unification},
    journal = {Phys. Lett. B},
    volume = {174},
    pages = {45},
    year = {1986},
    doi = {10.1016/0370-2693(86)91126-3}
}

@article{deGouvea2002,
    author        = {de Gouv\^ea, A. and Kayser, B. and Mohapatra, R. N.},
    title         = {Manifest {CP} violation from {M}ajorana phases},
    journal       = {Phys. Rev. D},
    volume        = {67},
    pages         = {053004},
    year          = {2003},
    doi           = {10.1103/PhysRevD.67.053004}
}

@article{Yang1967,
  author  = {Yang, C. N.},
  title   = {Some Exact Results for the Many-Body Problem in one Dimension with Repulsive Delta-Function Interaction},
  journal = {Phys. Rev. Lett.},
  volume  = {19},
  issue   = {23},
  pages   = {1312--1315},
  year    = {1967},
  doi     = {10.1103/PhysRevLett.19.1312}
}

@article{Baxter1972,
  author  = {Baxter, R. J.},
  title   = {Partition function of the eight-vertex lattice model},
  journal = {Ann. Phys.},
  volume  = {70},
  issue   = {1},
  pages   = {193--228},
  year    = {1972},
  doi     = {10.1016/0003-4916(72)90335-1}
}

@book{Baxter1982,
  author    = {Baxter, R. J.},
  title     = {Exactly Solved Models in Statistical Mechanics},
  publisher = {Academic Press},
  address   = {London, UK},
  year      = {1982},
  isbn      = {978-0-12-083180-7}
}

@article{Jimbo1989,
  author    = {Jimbo, M.},
  title     = {Introduction to the {Y}ang-{B}axter equation},
  journal   = {Int. J. Mod. Phys. A},
  volume    = {4},
  issue     = {15},
  pages     = {3759–3777},
  year      = {1989},
  doi       = {10.1142/S0217751X89001503}
}

@misc{Bilson2005,
  author    = {Bilson-Thompson, S. O.},
  title     = {A topological model of composite preons},
  year      = {2005},
  eprint    = {0503213},
  archivePrefix = {arXiv},
  primaryClass  = {hep-ph},
  doi           = {10.48550/arXiv.hep-ph/0503213}
}

@article{Artin1947,
  author    = {Artin, E.},
  title     = {Theory of Braids},
  journal   = {Ann. Math.},
  volume    = {48},
  issue     = {1},
  pages     = {101--126},
  year      = {1947},
  doi       = {10.2307/1969218}
}

@article{Faddeev1994,
  author        = {Faddeev, L. D. and Korchemsky, G. P.},
  title         = {High energy {QCD} as a completely integrable model},
  journal       = {Phys. Lett. B},
  volume        = {342},
  issue         = {1-4},
  pages         = {311--322},
  year          = {1995},
  doi           = {10.1016/0370-2693(94)01363-H}
}

@article{Nichita2012,
  author        = {Nichita, F.},
  title         = {Introduction to the {Y}ang-{B}axter Equation with Open Problems},
  journal       = {Axioms},
  volume        = {1},
  issue         = {1},
  pages         = {33-37},
  year          = {2012},
  doi           = {10.3390/axioms1010033}
}

@article{Arraut2019,
  author        = {Arraut, I.},
  title         = {The Quantum {Y}ang-{B}axter Conditions: The Fundamental Relations behind the {N}ambu-{G}oldstone Theorem},
  journal       = {Symmetry},
  volume        = {11},
  issue         = {6},
  pages         = {803},
  year          = {2019},
  doi           = {10.3390/sym11060803}
}

@article{Arraut2026,
  author  = {Arraut, I.},
  title   = {Predictions of the neutrino oscillations parameters}, 
  journal = {Int. J. Mod. Phys. A},
  volume  = {41},
  pages   = {2650079},
  year    = {2026},
  doi     = {10.1142/S0217751X2650079X}
}

@article{Wolfenstein1981,
  author    = {Wolfenstein, L.},
  title     = {{CP} properties of {M}ajorana neutrinos and double beta decay},
  journal   = {Phys. Lett. B},
  volume    = {107},
  issue    = {1--2},
  pages     = {77--79},
  year      = {1981},
  doi       = {10.1016/0370-2693(81)91151-5}
}

@article{Harrison2002_2,
    author        = "Harrison, P. F. and Scott, W. G.",
    title         = "{$\mu$ - $\tau$ reflection symmetry in neutrino mixing}",
    doi           = "10.1016/S0370-2693(02)02772-7",
    journal       = "Phys. Lett. B",
    volume        = "547",
    pages         = "219--228",
    year          = "2002"
}

@article{Grimus2004,
    author        = "Grimus, W. and Lavoura, L.",
    title         = "{A Nonstandard {CP} transformation leading to maximal atmospheric neutrino mixing}",
    doi           = "10.1016/j.physletb.2003.10.075",
    journal       = "Phys. Lett. B",
    volume        = "579",
    pages         = "113--122",
    year          = "2004"
}

@article{DayaBay2012,
    author = {An, F. P. and others},
    collaboration = {Daya Bay},
    title = {Observation of Electron-Antineutrino Disappearance at {D}aya {B}ay},
    journal = {Phys. Rev. Lett.},
    volume = {108},
    pages = {171803},
    year = {2012},
    doi = {10.1103/PhysRevLett.108.171803}
}

@article{DayaBay2022,
    author = {An, F. P. and others},
    collaboration = {Daya Bay},
    title = {Precision Measurement of Neutrino Oscillation Parameters with the Full {D}aya {B}ay Data Set},
    journal = {Phys. Rev. Lett.},
    volume = {130},
    number = {16},
    pages = {161802},
    year = {2023},
    doi = {10.1103/PhysRevLett.130.161802}
}

@article{Jarlskog2005,
    author = {Jarlskog, C.},
    title = {Invariants of lepton mass matrices and {CP} and {T} violation in neutrino oscillations},
    journal = {Phys. Lett. B},
    volume = {609},
    issue = {3-4},
    pages = {323-329},
    year = {2005},
    doi = {10.1016/j.physletb.2005.01.057}
}

@article{nufit2024,
  author  = {Esteban, I. and Gonzalez-Garcia, M. C. and Maltoni, M. and Martinez-Soler, I. and Pinheiro, J. P. and Schwetz, T.},
  title   = {{NuFit-6.0}: Updated global analysis of three-flavor neutrino oscillations},
  journal = {J. High Energ. Phys. $\textbf{2024}$},
  pages   = {216},
  year    = {{2024}},
  doi     = {10.1007/JHEP12(2024)216}
}

@misc{NuFit61,
    author = "Esteban, I. and Gonzalez-Garcia, M. C. and Maltoni, M. and Martinez-Soler, I. and Pinheiro, J. P. and Schwetz, T.",
    title = "{Nu{F}it 6.1 (2025), Three-neutrino fit based on data available in November 2025}",
    howpublished = "\url{http://www.nu-fit.org/node/309}",
    year = "2025",
    note = "www.nu-fit.org"
}

@article{Jones1987,
  author    = {Jones, V. F. R.},
  title     = {Hecke algebra representations of braid groups and link polynomials},
  journal   = {Ann. Math.},
  volume    = {126},
  issue    = {2},
  pages     = {335--388},
  year      = {1987},
  doi       = {10.2307/1971403}
}

@article{Jones1990,
  author   = {Jones, V. F. R.},
  title    = {Baxterization},
  journal  = {Int. J. Mod. Phys. B},
  volume   = {4},
  issue   = {5},
  pages    = {701--713},
  year     = {1990},
  doi      = {10.1142/S021797929000036X},
  note     = {[Proceedings of the conference on Yang--Baxter equations, conformal invariance and integrability in statistical mechanics and field theory (Canberra, Australia, 10--14 July 1989)]}
}

@article{Schechter1980,
  author    = {Schechter, J. and Valle, J. W. F.},
  title     = {Neutrino masses in {SU}(2) $\otimes$ {U}(1) theories},
  journal   = {Phys. Rev. D},
  volume    = {22},
  issue    = {9},
  pages     = {2227--2235},
  year      = {1980},
  doi       = {10.1103/PhysRevD.22.2227}
}

@article{Schechter1982,
  title     = {Neutrinoless double-beta decay in {SU}(2) $\otimes$ {U}(1) theories},
  author    ={Schechter, J. and Valle, J. W. F.},
  journal   = {Phys. Rev. D},
  volume    = {25},
  number    = {11},
  pages     = {2951},
  year      = {1982},
  doi       ={10.1103/PhysRevD.25.2951}
}

@article{Agostini2023,
    author = {Agostini, M. and others},
    title = {Toward the discovery of matter creation with neutrinoless double-beta decay},
    journal = {Rev. Mod. Phys.},
    volume = {95},
    number = {2},
    pages = {025002},
    year = {2023},
    doi = {10.1103/RevModPhys.95.025002}
}

@article{NOvA2026,
  title = {Precision Measurement of Neutrino Oscillation Parameters with 10 Years of Data from the {NO}v{A} Experiment},
  author = {Abubakar, S. and others},
  collaboration = {NOvA Collaboration},
  journal = {Phys. Rev. Lett.},
  volume = {136},
  issue = {1},
  pages = {011802},
  year = {2026},
  doi = {10.1103/x53y-2b86}
}

@article{JUNO2015,
    author        = {An, F. and others},
    collaboration = {JUNO Collaboration},
    title         = {Neutrino {P}hysics with {JUNO}},
    journal       = {J. Phys. G},
    volume        = {43},
    number        = {3},
    pages         = {030401},
    year          = {2016},
    doi           = {10.1088/0954-3899/43/3/030401}
}

@article{JUNO2022,
    author = {Abusleme, A. and others},
    collaboration = {JUNO Collaboration},
    title = {Sub-percent Precision Measurement of Neutrino Oscillation Parameters with {JUNO}},
    doi = {10.1088/1674-1137/ac8bc9},
    journal = {Chin. Phys. C},
    volume = {46},
    number = {12},
    pages = {123001},
    year = {2022}
}

@misc{NOvA_JUNO_2026,
  author        = {Abubakar, S. and others},
  collaboration = {NOvA Collaboration},
  title         = {Determining Neutrino Mass Ordering with {NO}v{A} and Upcoming {JUNO} Measurements},
  reportNumber  = {FERMILAB-PUB-26-0387-PPD},
  year          = {2026},
  eprint        = {2606.14121},
  archivePrefix = {arXiv},
  primaryClass  = {hep-ex}
}

@article{JUNO2026,
  author        = {Abusleme, A. and others },
  collaboration = {JUNO Collaboration},
  title         = {Measurement of reactor neutrino oscillation with the first {JUNO} data},
  journal       = {Nature},
  volume        = {654},
  pages         = {343--348},
  year          = {2026},
  doi           = {10.1038/s41586-026-10538-z}
}

@article{Ding2012,
  title = {Golden ratio neutrino mixing and {A}$_5$ flavor symmetry},
  author = {Ding, G. J. and Everett, L. L. and Stuart, A. J.},
  journal = {Nucl. Phys. B},
  volume = {857},
  issue = {2},
  pages = {219--253},
  year = {2012},
  doi = {10.1016/j.nuclphysb.2011.12.004}
}

@article{King2013,
  title = {Neutrino mass and mixing with discrete symmetry},
  author = {King, S. F. and Luhn, C.},
  journal = {Rep. Prog. Phys.},
  volume = {76},
  issue = {5},
  pages = {056201},
  year = {2013},
  doi = {10.1088/0034-4885/76/5/056201}
}

@article{King2013_2,
    author        = "King, S. F. and Merle, A. and Stuart, A. J.",
    title         = "{The Power of Neutrino Mass Sum Rules for Neutrinoless Double Beta Decay Experiments}",
    doi           = "10.1007/JHEP12(2013)005",
    journal       = "JHEP",
    volume        = "12",
    pages         = "005",
    year          = "2013"
}

@article{Nayak2008,
  title = {Non-Abelian anyons and topological quantum computation},
  author = {Nayak, C. and Simon, S. H. and Stern, A. and Freedman, M. and Das Sarma, S.},
  journal = {Rev. Mod. Phys.},
  volume = {80},
  issue = {3},
  pages = {1083--1159},
  numpages = {77},
  year = {2008},
  doi = {10.1103/RevModPhys.80.1083}
}

@article{Everett2009,
  title = {Icosahedral ({A}$_5$) family symmetry and the golden ratio prediction for solar neutrino mixing},
  author = {Everett, L. L. and Stuart, A. J.},
  journal = {Phys. Rev. D},
  volume = {79},
  issue = {8},
  pages = {085005},
  year = {2009},
  doi = {10.1103/PhysRevD.79.085005}
}

@book{Humphreys1990,
  title     = {Reflection Groups and Coxeter Groups},
  author    = {Humphreys, J. E.},
  series    = {Cambridge Studies in Advanced Mathematics},
  number    = {29},
  year      = {1990},
  publisher = {Cambridge University Press},
  address   = {Cambridge},
  doi       = {10.1017/CBO9780511623646}
}

@article{Casas2000,
  author        = {Casas, J. A. and Espinosa, J. R. and Ibarra, A. and Navarro, I.},
  title         = {General {RG} equations for physical neutrino parameters and their phenomenological implications},
  journal       = {Nucl. Phys. B},
  volume        = {573},
  issue        = {1-2},
  pages         = {652--684},
  year          = {2000},
  doi           = {10.1016/S0550-3213(99)00781-6}
}

@article{Antusch2002,
    author = {Antusch, S. and Kersten, J. and Lindner, M. and Ratz, M.},
    title = {Neutrino mass matrix running for nondegenerate seesaw scales},
    journal = {Phys. Lett. B},
    volume = {538},
    pages = {87--95},
    year = {2002},
    doi = {10.1016/S0370-2693(02)01960-3}
}

@article{Antusch2005,
    author = {Antusch, S. and Kersten, J. and Lindner, M. and Ratz, M. and Schmidt, M. A.},
    title = {Running neutrino mass parameters in see-saw scenarios},
    reportNumber = {TUM-HEP-576-05, DESY-05-013, SHEP-0504},
    doi = {10.1088/1126-6708/2005/03/024},
    journal = {J. High Energ. Phys.},
    volume = {03},
    pages = {024},
    year = {2005}
}

@article{Ohlsson2014,
    author        = {Ohlsson, T. and Zhou, S.},
    title         = {Renormalization group running of neutrino parameters},
    reportNumber  = {KTH-TH-13-17},
    doi           = {10.1038/ncomms6153},
    journal       = {Nature Commun.},
    volume        = {5},
    pages         = {5153},
    year          = {2014}
}

@article{Feruglio2011,
    author        = {Feruglio, F. and Paris, A.},
    title         = {The Golden Ratio Prediction for the Solar Angle from a Natural Model with ${A}_5$ Flavour Symmetry},
    journal       = {J. High Energ. Phys.},
    volume        = {2011},
    number        = {3},
    pages         = {101},
    year          = {2011},
    doi           = {10.1007/JHEP03(2011)101}
}

@article{Feruglio2013,
  author        = {Feruglio, F. and Hagedorn, C. and Ziegler, R.},
  title         = {Lepton mixing parameters from discrete and {CP} symmetries},
  journal       = {J. High Energ. Phys.},
  volume        = {2013},
  number        = {7},
  pages         = {027},
  year          = {2013},
  doi           = {10.1007/JHEP07(2013)027}
}

@book{Chari1994,
  author    = {Chari, V. and Pressley, A.},
  title     = {A Guide to Quantum Groups},
  publisher = {Cambridge University Press},
  address   = {Cambridge},
  year      = {1994},
  isbn      = {978-0-521-43305-1}
}

@article{Temperley1971,
  author    = {Temperley, H. N. V. and Lieb, E. H.},
  title     = {Relations between the 'percolation' and 'colouring' problem and other graph-theoretical problems associated with regular planar lattices},
  journal   = {Proceedings of the Royal Society of London. Series A, Mathematical and Physical Sciences},
  volume    = {322},
  issue     = {1549},
  pages     = {251--280},
  year      = {1971},
  doi       = {10.1098/rspa.1971.0067}
}

@article{Iwahori1964,
  author    = {Iwahori, N.},
  title     = {On the Structure of a {H}ecke Ring of a {C}hevalley Group over a Finite Field},
  journal   = {Journal of the Faculty of Science, University of Tokyo},
  volume    = {10},
  issue     = {2},
  pages     = {215--236},
  year      = {1964},
  publisher = {University of Tokyo},
  doi       = {10.15083/00039900}
}

@article{Formaggio2021,
    author = {Formaggio, J. A. and de Gouv\^ea, A. L. C. and Robertson, R. G. H.},
    title = {Direct measurements of neutrino mass},
    journal = {Phys. Rept.},
    volume = {914},
    pages = {1--54},
    year = {2021},
    doi = "10.1016/j.physrep.2021.02.002"
}

@article{Kitaev2003,
  author    = {Kitaev, A. Y.},
  title     = {Fault-tolerant quantum computation by anyons},
  journal   = {Ann. Phys.},
  volume    = {303},
  issue     = {1},
  pages     = {2--30},
  year      = {2003},
  doi       = {10.1016/S0003-4916(02)00018-0}
}

@article{PDG2024,
    author        = {Navas, S. and others},
    collaboration = {Particle Data Group},
    title         = {Review of Particle Physics},
    journal       = {Phys. Rev. D},
    volume        = {110},
    pages         = {030001},
    year          = {2024},
    doi           = {10.1103/PhysRevD.110.030001}
}

@article{T2K2025,
    author        = {Abe, A. and others},
    collaboration = {T2K Collaboration},
    title         = {Testing {T2K}'s Bayesian constraints with priors in alternate parameterisations},
    journal       = {Eur. Phys. J. C},
    volume        = {85},
    pages         = {1414},
    year          = {2025},
    doi           = {10.1140/epjc/s10052-025-14836-0}
}

@article{NOvAT2K2025,
    author        = "{The NOvA Collaboration} and {The T2K Collaboration}",
    title         = {Joint neutrino oscillation analysis from the {T2K} and {NO}v{A} experiments},
    journal       = {Nature},
    volume        = {646},
    pages         = {818--824},
    year          = {2025},
    doi           = {10.1038/s41586-025-09599-3}
}

@article{Rodejohann2011,
    author        = {Rodejohann, W.},
    title         = {Neutrino-less double beta decay and particle physics},
    journal       = {Int. J. Mod. Phys. E},
    volume        = {20},
    pages         = {1833--1930},
    year          = {2011},
    doi           = {10.1142/S0218301311020186}
}

@article{Rodejohann2012,
    author        = {Rodejohann, W.},
    title         = {Neutrinoless double-beta decay and neutrino physics},
    journal       = {J. Phys. G},
    volume        = {39},
    number        = {12},
    pages         = {124008},
    year          = {2012},
    doi           = {10.1088/0954-3899/39/12/124008}
}

@article{Chen2009,
    author        = {Chen, M.-C. and Mahanthappa, K. T.},
    title         = {Group Theoretical Origin of {CP} Violation},
    journal       = {Phys. Lett. B},
    volume        = {681},
    pages         = {444--447},
    year          = {2009},
    doi           = {10.1016/j.physletb.2009.10.059}
}

@article{Chen2014,
    author        = {Chen, M.-C. and Fallbacher, M. and Mahanthappa, K. T. and Ratz, M. and Trautner, A.},
    title         = {{CP} Violation from Finite Groups},
    journal       = {Nucl. Phys. B},
    volume        = {883},
    pages         = {267--305},
    year          = {2014},
    doi           = {10.1016/j.nuclphysb.2014.03.023}
}

@misc{Hyper-Kamiokande2018,
    author        = {Abe, K. and others},
    collaboration = {Hyper-Kamiokande Collaboration},
    title         = {Hyper-{K}amiokande {D}esign {R}eport},
    eprint        = {1805.04163},
    archivePrefix = {arXiv},
    primaryClass  = {physics.ins-det},
    journal       = {KEK Preprint},
    volume        = {2016-21},
    year          = {2018},
    url           = {https://arxiv.org/abs/1805.04163}
}

@article{DUNE2020,
    author        = {Abi, B. and others},
    collaboration = {DUNE Collaboration},
    title         = {Deep {U}nderground {N}eutrino {E}xperiment ({DUNE}), {T}echnical {D}esign {R}eport, {V}olume {II}: {DUNE} {P}hysics},
    journal       = {Eur. Phys. J. C},
    volume        = {80},
    number        = {10},
    pages         = {978},
    year          = {2020},
    doi           = {10.1140/epjc/s10052-020-08467-y}    
}

@article{KamLAND_Zen2022,
    author = {Abe, S. and others},
    collaboration = {KamLAND-Zen},
    title = {First Search for the Majorana Nature of Neutrinos in the Inverted Mass Ordering Region with {K}am{LAND}-{Z}en},
    journal = {Phys. Rev. Lett.},
    volume = {130},
    number = {5},
    pages = {051801},
    year = {2023},
    doi = {10.1103/PhysRevLett.130.051801}
}

@article{Banchoff1967,
  author    = {Banchoff, T. F.},
  title     = {Critical points and curvature for embedded polyhedra},
  journal   = {J. Differential Geom.},
  volume    = {1},
  number    = {3-4},
  pages     = {257--268},
  year      = {1967},
  doi       = {10.4310/jdg/1214428092}
}

@inbook{Coxeter1973,
  author     = {Coxeter, H. S. M.},
  title      = {Regular Polytopes},
  chapter    = {2.5},
  edition    = {3rd},
  publisher  = {Dover Publications},
  address    = {New York},
  year       = {1973},
  series     = {Dover Books on Mathematics},
  isbn       = {978-0-486-61480-9},
  isbn-10    = {0-486-61480-8},
  lccn       = {73-084364},
  url        = {https://store.doverpublications.com/products/9780486614809}
}

@article{Appelquist1975,
  author   = {Appelquist, T. W. and Carazzone, J. J.},
  title    = {Infrared singularities and massive fields},
  journal  = {Phys. Rev. D},
  volume   = {11},
  number   = {10},
  pages    = {2856--2861},
  year     = {1975},
  doi      = {10.1103/PhysRevD.11.2856}
}

@article{Lesgourgues2006,
  author        = {Lesgourgues, J. and Pastor, S.},
  title         = {Massive neutrinos and cosmology},
  journal       = {Physics Reports},
  volume        = {429},
  number        = {6},
  pages         = {307--379},
  year          = {2006},
  doi           = {10.1016/j.physrep.2006.04.001}
}

@article{Pasquier1990,
  title={Common structures between finite systems and conformal field theories through quantum groups},
  author={Pasquier, V. and Saleur, H.},
  journal={Nucl. Phys. B},
  volume={330},
  number={2-3},
  pages={523--556},
  year={1990},
  doi={10.1016/0550-3213(90)90122-T}
}

@article{Pakvasa2008,
  author    = {Pakvasa, S. and Rodejohann, W. and Weiler, T. J.},
  title     = {Flavor Ratios of Astrophysical Neutrinos: Implications for Precision Measurements},
  journal   = {J. High Energ. Phys.},
  volume    = {02},
  pages     = {005},
  year      = {2008},
  doi       = {10.1088/1126-6708/2008/02/005}
}

@article{Bustamante2015,
  author    = {Bustamante, M. and Beacom, J. F. and Winter, W.},
  title     = {Theoretically Palatable Flavor Combinations of Astrophysical Neutrinos},
  journal   = {Phys. Rev. Lett.},
  volume    = {115},
  issue     = {16},
  pages     = {161302},
  year      = {2015},
  doi       = {10.1103/PhysRevLett.115.161302}
}

@article{Li2015,
    author = {Li, C.-C. and Ding, G.-J.},
    title = {Lepton mixing in $A_5$ family symmetry and generalized {CP}},
    doi = {10.1007/JHEP05(2015)100},
    journal = {J. High Energ. Phys.},
    volume = {05},
    pages = {100},
    year = {2015}
}

@article{DiIura2015,
    author        = {Di Iura, A. and Hagedorn, C. and Meloni, D.},
    title         = {Lepton mixing from the interplay of the alternating group $A_5$ and {CP}},
    doi           = {10.1007/JHEP08(2015)037},
    journal       = {J. High Energ. Phys.},
    volume        = {08},
    pages         = {037},
    year          = {2015}
}

@article{Ballett2015,
    author        = {Ballett, P. and Pascoli, S. and Turner, J.},
    title         = {Mixing angle and phase correlations from $A_5$ with generalized {CP} and their prospects for discovery},
    doi           = {10.1103/PhysRevD.92.093008},
    journal       = {Phys. Rev. D},
    volume        = {92},
    number        = {9},
    pages         = {093008},
    year          = {2015}
}

@article{Xing2016,
    author        = {Xing, Z.-Z. and Zhao, Z.-H.},
    title         = {A review of $\mu-\tau$ flavor symmetry in neutrino physics},
    doi           = {10.1088/0034-4885/79/7/076201},
    journal       = {Rep. Prog. Phys.},
    volume        = {79},
    number        = {7},
    pages         = {076201},
    year          = {2016}
}

@article{Barry2011,
    author        = {Barry, J. and Rodejohann, W.},
    title         = {Neutrino Mass Sum-rules in Flavor Symmetry Models},
    doi           = {10.1016/j.nuclphysb.2010.08.015},
    journal       = {Nucl. Phys. B},
    volume        = {842},
    pages         = {33--50},
    year          = {2011}
}

@article{Gehrlein2016,
    author        = {Gehrlein, J. and Merle, A. and Spinrath, M.},
    title         = {Predictivity of Neutrino Mass Sum Rules},
    doi           = {10.1103/PhysRevD.94.093003},
    journal       = {Phys. Rev. D},
    volume        = {94},
    number        = {9},
    pages         = {093003},
    year          = {2016}
}

@inbook{Goodman1989,
    author    = {Goodman, F. M. and de la Harpe, P. and Jones, V. F. R.},
    title     = {Coxeter Graphs and Towers of Algebras},
    series    = {Mathematical Sciences Research Institute Publications},
    chapter   = {2.7--2.9},
    volume    = {14},
    publisher = {Springer-Verlag},
    address   = {New York},
    year      = {1989},
    isbn      = {0-387-96979-9},
    doi       = {10.1007/978-1-4613-9641-3},
    note       = {\href{https://link.springer.com/book/10.1007/978-1-4613-9641-3}{Springer--Verlag}}
}

@misc{LEGEND2021,
    author        = {Abgrall, N. and others},
    collaboration = {LEGEND Collaboration},
    title         = {The {L}arge {E}nriched {G}ermanium {E}xperiment for {N}eutrinoless $\beta\beta$ {D}ecay: {LEGEND}-1000 Preconceptual Design Report},
    eprint        = {2107.11462},
    archivePrefix = {arXiv},
    primaryClass  = {physics.ins-det},
    month         = {7},
    year          = {2021}
}

@article{CUPID2025,
    author        = {Alfonso, K. and others},
    collaboration = {CUPID Collaboration},
    title         = {{CUPID}, the {CUORE} upgrade with particle identification},
    doi           = {10.1140/epjc/s10052-025-14352-1},
    journal       = {Eur. Phys. J. C},
    volume        = {85},
    number        = {7},
    pages         = {737},
    year          = {2025}
}

@article{nEXO2021,
    author        = {Adhikari, G. and others},
    collaboration = {nEXO Collaboration},
    title         = {n{EXO}: neutrinoless double beta decay search beyond $10^{28}$ year half-life sensitivity},
    doi           = {10.1088/1361-6471/ac3631},
    journal       = {J. Phys. G},
    volume        = {49},
    number        = {1},
    pages         = {015104},
    year          = {2022}
}

@article{Capozzi2026,
    author        = {Capozzi, F. and Lisi, E. and Marcone, F. and Marrone, A. and Palazzo, A.},
    title         = {Updated bounds on the (1,2) neutrino oscillation parameters after first {JUNO} results},
    doi           = {10.1103/cxqw-1bty},
    journal       = {Phys. Rev. D},
    volume        = {114},
    number        = {1},
    pages         = {016026},
    year          = {2026}
}

@article{He2026,
    author        = {He, X.-G.},
    title         = {Modified tri-bimaximal neutrino mixing confronted by {JUNO} $\theta_{12}$ measurement},
    doi           = {10.1016/j.physletb.2026.140270},
    journal       = {Phys. Lett. B},
    volume        = {874},
    pages         = {140270},
    year          = {2026}
}

@article{Ding2026,
    author        = {Ding, G.-J. and Li, C.-C. and Lu, J.-N. and Petcov, S. T.},
    title         = {Discrete flavour and {CP} symmetries in light of {JUNO} and neutrino global fit},
    doi           = {10.1007/JHEP07(2026)256},
    journal       = {J. High Energ. Phys.},
    volume        = {07},
    pages         = {256},
    year          = {2026}
}

@article{Petcov2026,
    author        = {Petcov, S. T. and Titov, A. V.},
    title         = {Viability of $A_4$, $S_4$ and $A_5$ Flavour Symmetries in Light of the First {JUNO} Result},
    doi           = {10.1016/j.physletb.2026.140295},
    journal       = {Phys. Lett. B},
    volume        = {874},
    pages         = {140295},
    year          = {2026}
}

@article{ArkaniHamed2001,
    author = {Arkani-Hamed, N. and Cohen, A. G. and Georgi, H.},
    title = {(De)constructing dimensions},
    journal = {Phys. Rev. Lett.},
    volume = {86},
    pages = {4757--4761},
    year = {2001},
    doi = "10.1103/PhysRevLett.86.4757"
}

@article{Giudice2017,
    author = {Giudice, G. F. and McCullough, M.},
    title = {A Clockwork Theory},
    reportNumber = {CERN-TH-2016-224},
    journal = {J. High Energ. Phys.},
    volume = {02},
    pages = {036},
    year = {2017},
    doi = {10.1007/JHEP02(2017)036}
}

@article{IceCubeGen2_2020,
  author        = {Aartsen, M. G. and others},
  collaboration = {IceCube-Gen2 Collaboration},
  title         = {{Ice{C}ube-{G}en2: the window to the extreme Universe}},
  journal       = {J. Phys. G},
  volume        = {48},
  number        = {6},
  pages         = {060501},
  year          = {2021},
  doi           = {10.1088/1361-6471/abbd48}
}

@article{KM3Net2016,
  author        = {Adri{\'a}n-Mart{\'i}nez, S. and others},
  collaboration = {KM3NeT Collaboration},
  title         = {{Letter of intent for KM3NeT 2.0}},
  journal       = {J. Phys. G},
  volume        = {43},
  number        = {8},
  pages         = {084001},
  year          = {2016},
  doi           = {10.1088/0954-3899/43/8/084001}
}

@article{Jenkins2008,
  author        = {Jenkins, E. E. and Manohar, A. V.},
  title         = {Tribimaximal mixing, leptogenesis, and $\theta_{13}$},
  journal       = {Phys. Lett. B},
  volume        = {668},
  pages         = {210--215},
  year          = {2008},
  doi           = {10.1016/j.physletb.2008.08.028}
}

@article{Bertuzzo2009,
  author        = {Bertuzzo, E. and Di Bari, P. and Feruglio, F. and Nardi, E.},
  title         = {Flavor symmetries, leptogenesis and the absolute neutrino mass scale},
  journal       = {J. High Energy Phys.},
  volume        = {11},
  pages         = {036},
  year          = {2009},
  doi           = {10.1088/1126-6708/2009/11/036},
  eprint        = {0908.0161},
  archivePrefix = {arXiv},
  primaryClass  = {hep-ph}
}

\end{document}